\documentclass[11pt,letterpaper]{article}

\pdfoutput=1
\usepackage{jcappub}

\usepackage{framed}
\usepackage{appendix}
\usepackage{array}
\newcolumntype{x}[1]{>{\centering\arraybackslash}p{#1}}
\usepackage{bm}
\usepackage{color}
\usepackage{graphicx}
\usepackage{cancel}
\usepackage{amsfonts}
\usepackage{amssymb}
\usepackage{amsmath}
\usepackage{calc}
\usepackage{dcolumn}
\usepackage{mathrsfs}
\usepackage{mathtools}
\usepackage{soul}
\usepackage{amsmath}
\usepackage[normalem]{ulem}

\usepackage[normalem]{ulem} 

\newcommand{\eg}{e.g.~}
\newcommand{\ie}{i.e.~}

\newcommand{\Eq}[1]{Eq.~\eqref{#1}}
\newcommand{\Fig}[1]{Fig.~\ref{#1}}
\newcommand{\Sec}[1]{Sec.~\ref{#1}}

\newcommand{\Ref}[1]{Ref.~\cite{#1}}           

\newcommand{\Hcur}{\mathcal{H}}

\newcommand{\beq}{\begin{equation}}
\newcommand{\eeq}{\end{equation}}

\newcommand{\ud}{\text{d}}
\newcommand{\bol}[1]{\boldsymbol{#1}}

\newcommand{\ER}{E_\text{R}}
\newcommand{\Ed}{E'}

\newcommand{\vmin}{v_\text{min}}

\newcommand{\teta}{\tilde{\eta}}

\newcommand{\ignore}[1]{}

\definecolor{rossoCP3}{cmyk}{0,.88,.77,.40}
\definecolor{verdeCP3}{rgb}{0.09765625, 0.57421875, 0.1015625}
\definecolor{bluCP3}{rgb}{0, 0.23, 0.67}

\ifx\pdfoutput\undefined
\usepackage[dvips,bookmarks]{hyperref}    
\else
\usepackage{hyperref}    
\fi
\hypersetup{colorlinks, bookmarksopen, bookmarksnumbered,
citecolor=verdeCP3, linkcolor=bluCP3, pdfstartview=FitH, urlcolor=rossoCP3}

\notoc

\newcommand{\AddrUCLA}{Department of Physics and Astronomy, UCLA, 475 Portola Plaza, Los Angeles, CA 90095 (USA)}

\newcommand{\AddrCERN}{Theory Department, CERN, CH-1211 Geneva 23, Switzerland}

\newcommand{\AddrIFIC}{Instituto de F\'{i}sica Corpuscular (IFIC), CSIC-Universitat de Val\`{e}ncia, Apartado de Correos 22085, E-46071 Valencia, Spain}

\begin{document}

\title{Unified Halo-Independent Formalism From Convex Hulls for Direct Dark Matter Searches}

\subheader{CERN-TH-2017-159}

\author[a]{Graciela B.~Gelmini,}
\author[b]{Ji-Haeng Huh,}
\author[a,c]{Samuel J.~Witte}

\affiliation[a]{\AddrUCLA}
\affiliation[b]{\AddrCERN}
\affiliation[c]{\AddrIFIC}

\emailAdd{gelmini@physics.ucla.edu}
\emailAdd{ji-haeng.huh@cern.ch}
\emailAdd{sam.witte@ific.uv.es}

\abstract{
Using the Fenchel-Eggleston theorem for convex hulls (an extension of the Caratheodory theorem), we prove that  any likelihood can be maximized by either a dark matter 1- speed distribution $F(v)$ in Earth's frame or 2- Galactic velocity distribution $f^{\rm gal}(\vec{u})$, consisting of a sum of delta functions. The former case applies only to time-averaged rate measurements and  the maximum number of delta functions is  $({\mathcal N}-1)$, where ${\mathcal N}$ is the total number of data entries. The second case applies to any harmonic expansion coefficient of the time-dependent rate and the maximum number of terms is ${\mathcal N}$. Using time-averaged rates, the aforementioned form of $F(v)$ results in a  piecewise constant unmodulated halo function $\tilde\eta^0_{BF}(\vmin)$ (which is an integral of the speed distribution) with at most $({\mathcal N}-1)$ downward steps. The authors had previously proven this result for likelihoods comprised of at least one extended likelihood, and found the best-fit halo function to be unique. This uniqueness, however, cannot be guaranteed in the more general analysis applied to arbitrary likelihoods. Thus we introduce a method for determining whether there exists a unique best-fit halo function, and provide a procedure for constructing either a pointwise confidence band, if the best-fit halo function is unique, or a degeneracy band, if it is not. Using measurements of modulation amplitudes, the aforementioned form of $f^{\rm gal}(\vec{u})$, which is a sum of Galactic streams, yields a periodic 
time-dependent halo function $\tilde\eta_{BF}(\vmin, t)$ which at any fixed time is a piecewise constant function of $\vmin$ with at most ${\mathcal N}$ downward steps.  In this case, we explain how to construct pointwise confidence and degeneracy bands from the time-averaged halo function. Finally, we show that requiring an isotropic Galactic velocity distribution leads to a Galactic speed distribution $F(u)$ that is once again a sum of delta functions, and produces a time-dependent $\tilde\eta_{BF}(\vmin, t)$  function (and a time-averaged $\tilde\eta^0_{BF}(\vmin)$) that is piecewise linear, differing significantly from best-fit halo functions obtained without the assumption of isotropy. }


\maketitle

\flushbottom

\newpage


\section{Introduction}

The non-gravitational interactions of the most dominant form of matter in the Universe, the dark matter (DM), remain elusive to-date.
Weakly interacting massive particles (WIMP's) are among the most studied DM particle candidates, in part  because they commonly appear in extensions of the Standard Model. 

WIMPs are being searched for in different ways. Direct DM detection experiments attempt to measure the recoil energy of nuclei after they collide with WIMPs in the Galactic dark halo.  The current status of DM direct detection experiments remains ambiguous, with a few experiments having claimed  a potential DM signal~\cite{Bernabei:2010mq, Aalseth:2010vx, Aalseth:2012if, Aalseth:2011wp, Aalseth:2014eft, Aalseth:2014jpa, Agnese:2013rvf} and all others reporting upper bounds, some of which appear to be in irreconcilable conflict with the putative detection claims for nearly all particle candidates~\cite{Angle:2011th, Aprile:2011hi, Aprile:2012nq, Felizardo:2011uw, Archambault:2012pm, Behnke:2012ys, Ahmed:2012vq, Agnese:2015ywx,Akerib:2015rjg,Agnese:2015nto,Agnese:2014aze,Amole:2015lsj,Aprile:2016swn,Akerib:2016vxi,Tan:2016zwf,Amole:2017dex,Geng:2017ypy,Gelmini:2017aad,Aprile:2017yea,Aprile:2017kek,Aprile:2017aas,Aprile:2017iyp,Akerib:2017kat,Agnese:2017jvy}

There are three main elements of the predicted rate in any direct DM experiment:  the detector response, the DM particle  model, and the local characteristics of  the dark halo model. The uncertainties associated with these inputs can significantly affect the expected recoil spectrum (both in shape and magnitude) for a particular experiment, as well as the observed compatibility between different experimental data. 

There are two different methods for comparing the results of different direct DM experiments.  In the halo-dependent method, used since the inception of direct detection in the 1980's,  it is necessary to model the local DM density and velocity distribution. But there are large uncertainties in the local characteristics of the Galactic halo, and in trying to circumvent them, a halo-independent data comparison method has recently been developed in which no model for the dark halo is assumed. These methods infer properties of the local halo from different direct detection data sets for a fixed DM particle model, and then compare these inferred halo characteristics to determine the compatibility of the data sets.  Finding that different putative signals are compatible under the assumption of some particular DM particle models and not others, in a halo-independent manner that avoids all astrophysical uncertainties, would constitute a clear indication of the characteristics of the DM that could produce the signals \footnote{Our halo-independent approach differs from others, sometimes called (halo) model-independent, which try to mitigate the impact of astrophysical uncertainties on the reconstruction of particle physics properties by assuming a particular parameterization of the DM velocity or speed distribution  (\eg taking the log of the speed distribution to be an expansion in Legendre or Chebyshev polynomails), see \eg~\cite{Peter:2011eu,Kavanagh:2013wba,Kavanagh:2014rya}. In these methods the halo and the particle model parameters are fit together using the data. We instead do not assume any functional parameterization of the DM speed or velocity distribution (its form is imposed by the theorems we use), and we do not fit DM particle parameters (we assume a particular DM model for each data analysis we perform).}.

In the  halo-independent data analysis, putative measurements and bounds on the nuclear recoil rate are translated into measurements and bounds on  a function common to all direct detection experiments that we call the halo function $\teta(\vmin,t)$. This  function, up to a multiplicative constant, is  the average of the inverse speed over all DM particle speeds larger than some minimum value $\vmin$. It is a periodic function of time due to Earth's rotation around the Sun.  This periodic function contains all of the dependence of the predicted rate on the halo model (see \eg\cite{Fox:2010bz,Fox:2010bu,Frandsen:2011gi,Gondolo:2012rs,HerreroGarcia:2012fu,Frandsen:2013cna,DelNobile:2013cta,Bozorgnia:2013hsa,DelNobile:2013cva,DelNobile:2013gba,DelNobile:2014eta,Feldstein:2014gza,Fox:2014kua,Gelmini:2014psa,Cherry:2014wia,DelNobile:2014sja,Scopel:2014kba,Feldstein:2014ufa,Bozorgnia:2014gsa,Blennow:2015oea,DelNobile:2015lxa,Anderson:2015xaa,Blennow:2015gta,Scopel:2015baa,Ferrer:2015bta,Wild:2016myz,Gondolo:2017jro,Witte:2017qsy}). 

Early halo-independent analyses were limited in the way they handled putative signals. Only weighted averages on $\vmin$ intervals of the unmodulated component of $\teta(\vmin,t)$, $\teta^0(\vmin)$, and of the amplitude of the annually modulated component, $\teta^1(\vmin)$, (see \Eq{eta} below) were plotted against upper bounds in $\vmin - \teta$ plane (see \eg\cite{Fox:2010bz,Frandsen:2011gi,Gondolo:2012rs,DelNobile:2013cva}). Comparing weighted averages with upper limits at a particular confidence level (CL) does not allow for a statistically meaningful assessment of compatibility or incompatibility of different data sets. It requires a by-eye interpretation of the compatibility of putative signals and with null searches (an in principle comparisons between different putative signals), and thus only qualitative and statistically ambiguous statements can be made regarding the viability of potential detections.

Recently,  halo-independent analyses have been developed that provide a clear statistical meaning to time-averaged measurements of the rate.  Using extended likelihoods Refs.~\cite{Fox:2014kua, Gelmini:2015voa} (or using global likelihoods containing at most an extended likelihood Ref.~\cite{Gelmini:2016pei}) have shown how to determine the unique best-fit $\tilde\eta^0_{BF}(\vmin)$ function and construct a two-sided pointwise confidence band in the $\vmin-\teta^0$ plane, at any chosen CL~\cite{Gelmini:2015voa, Gelmini:2016pei}. In this way, upper bounds at a particular CL can be compared to a confidence band at a particular CL to assess if they are compatible (see~\cite{Gelmini:2015voa} for a discussion). However, until now, this procedure was strongly limited by the fact that it could not be applied to exclusively binned data or to measurements of modulation amplitudes. In this paper we address these limitations. 

In this paper we use theorems that apply to convex hulls to generalize to arbitrary likelihoods the formalism for identifying the best-fit halo function and constructing two-sided pointwise confidence bands. This work thus presents a clear formalism that allows for unambiguous statistical statements regarding the compatibility of putative and null signals at any chosen confidence level, removing all caveats that limited the methods presented in~\cite{Fox:2014kua, Gelmini:2015voa,Gelmini:2016pei}.

The mathematical background we use is also the basis of the formalism in~\cite{Gondolo:2017jro}, and also similar to a lesser degree to that of~\cite{Ibarra:2017mzt},  although the language, methods and purposes of these papers are quite different (see Sec.~\ref{sec:theorems}).

In Sec.~\ref{haloindep} we review the halo-independent formalism previously developed in~\cite{Gelmini:2015voa, Gelmini:2016pei}, in which the predicted rate is written as the convolution of  a (DM particle model and detector dependent) response function and the speed distribution (readers familiar with~\cite{Gelmini:2015voa, Gelmini:2016pei} may wish to skip this section).

  In Sec.~\ref{Likelihood} we explicitly write the likelihoods, and show that they depend on the halo model only through the predicted rates.
In Sec.~\ref{sec:theorems}, we show that the predicted rates form the convex hull of the response functions, we  make use of  the Fenchel-Eggleston~\cite{zbMATH03084780,zbMATH03140980} theorem (an extension of the Caratheodory theorem~\cite{zbMATH02644208}) to write the time-averages rates in terms of a DM speed distribution function in Earth's frame $F(v)$ that consists of a sum of at most 
$(\mathcal{N} - 1)$ delta functions in speed, where $\mathcal{N}$ is the total number of data entries (see \Eq{eq:curlyN}).
Since the likelihoods can always be maximized for a particular set of predicted rates, the maximum of any likelihood can be always found with a speed distribution of this form.
This implies that  the time-averaged best-fit halo function $\teta^0_{BF}(\vmin)$, which is an integral over the speed distribution, is piecewise constant with at most $(\mathcal{N} - 1)$ downward steps.
This procedure reproduces the results previously obtained in~\cite{Gelmini:2015voa, Gelmini:2016pei} for analyses that contain at least one extended likelihood, and extends the result to any other likelihood depending only on unmodulated rates.  There is, however, one caveat: using  extended likelihoods the  best-fit $\teta$ is guaranteed to be unique, while for likelihoods depending only on binned data, \eg Poisson or Gaussian, the best-fit function may or may not be unique.
The method for identifying the best-fit halo function $\teta^0_{BF}$ and determining if it is unique is the subject of \Sec{sec:bf}.  

\Sec{sec:cband} contains a procedure for deriving pointwise confidence bands, if $\teta^0_{BF}$ is unique, and a method for identifying the parameter space spanned by all degenerate best-fit halo functions (which we refer to as {\it{degeneracy band}}), if $\teta^0_{BF}$ is not unique.
In \Sec{sec:examples} we provide examples of the aforementioned methods using mock data. In \Sec{sec:modulation} we show how to extend the formalism presented in the previous sections, based on the use of the Fenchel-Eggleston theorem, to all coefficients of a harmonic expansion of the time-dependent rate. We finish with a brief summary in \Sec{sec:summary}. 

\section{Our Halo-independent formalism \label{haloindep}}

In this section we introduce the necessary formalism  that will be used in the following sections. The primary purpose is to write the expected rate in a direct detection experiment as the convolution of a DM particle model and detector dependent response function and the DM velocity or speed distribution.

We begin with the calculation of the scattering rate for a generic DM candidate and a generic direct detection experiment. The differential recoil rate per unit of detector mass  as a function of nuclear recoil energy $\ER$  for WIMPs of mass $m$ scattering off a target nuclide $T$ of mass $m_T$, in a particular experiment is given by
\begin{equation}\label{diffrate}
\frac{\ud R_T}{\ud \ER} = \frac{\rho}{m}\frac{C_T}{m_T}\int_{v \geqslant \vmin(\ER)} \, \ud^3 \, v \, f(\bol{v},t) \, v \, \frac{\ud \sigma_T}{\ud \ER}(\ER, \bol{v}) \, ,
\end{equation}
where  $C_T$ is the mass fraction of the nuclide $T$ in a detector, thus $C_T/m_T$ is the number a target nuclides T unit of detector mass, $\rho$ and $f(\bol{v},t)$ are the local DM density velocity distribution in Earth's frame  and $\ud \sigma_T / \ud \ER$ is the WIMP-nuclide differential cross section in the lab frame. $f(\bol{v},t)$ is a function of time $t$ due to the
rotation of the Earth around the Sun. When the detector includes multiple target elements or nuclides $T$, the differential rate is a sum over all of them
\begin{equation}\label{sum_diffrate}
\frac{\ud R}{\ud \ER} = \sum_T \frac{\ud R_T}{\ud \ER} \, .
\end{equation}

In some particle models the DM particle scatters to a different  particle of mass $m^{\prime} = m + \delta$, so that the DM-nucleus scattering is inelastic. If  $\delta > 0$  the scattering is endothermic and if $\delta <0$ it is exothermic. In elastic scattering $\delta = 0$. With $|\delta| \ll m$ and $\mu_T |\delta|/m^2 \ll 1$, $\vmin(\ER)$ is given by 
\begin{equation}\label{eq:vmin}
\vmin(\ER) = \frac{1}{\sqrt{2 m_T \ER}} \left| \frac{m_T \ER}{\mu_T} +\delta \right| \, .
\end{equation}
Here $\mu_T$ is the reduced mass of the WIMP-nucleus system. From \Eq{eq:vmin} one obtains the range of possible recoil energies that can be imparted to a target nucleus by a DM particle traveling at speed $v$ in Earth's frame,  $\ER^{T,-}(v) < \ER< \ER^{T,+}(v)$, where
\begin{equation}\label{eq:Ebranch}
\ER^{T,\pm} (v) = \frac{\mu_T^2 v^2}{2 m_T} \left( 1 \pm \sqrt{1-\frac{2\delta}{\mu_T v^2}} \right)^2 \, .
\end{equation}
\Eq{eq:Ebranch} shows that for endothermic scattering, $\delta > 0$, there exists a nontrivial kinematic endpoint in DM speed
\begin{equation}\label{eq:vdelta}
v_\delta^T = \sqrt{2 \delta / \mu_T}\, ,
\end{equation}
such  a DM particle interacting with initial $v <v_\delta^T$ cannot induce a nuclear recoil (because there is not enough energy in the collision to produce the final DM particle with larger mass). When multiple target nuclides are present in a detector, we use $v_\delta$ to denote the smallest of all $v_\delta^T$  for all $T$ in the detector. For exothermic and elastic scattering $v_\delta=0$ and WIMPs reaching the detector with any speed can always produce a nuclear recoil when they interact within the detector.

Only for elastic scattering off a single  target nuclide is the relation between $\vmin$ and the recoil energy $\ER$ unique. In all other scenarios, one needs to choose wether to treat one or the other as the independent variable. If $\ER$ is considered the independent variable, then $\vmin$ is the minimum speed necessary for the incoming DM particle to impart a nuclear recoil $\ER$ to the target nucleus and, thus it depends on the target nuclide $T$ through its mass $m_T$, $\vmin^T=\vmin(\ER,m_T,m)$. This has been the more common approach in early halo-independent analysis papers (see \eg\cite{Fox:2010bz,Frandsen:2011gi,Frandsen:2013cna}). Alternatively, one can choose to treat $\vmin$ as the independent variable, in which case $\ER^{T}$ is understood to be the extremum recoil energy (the maximum for elastic scattering, and either the maximum or the minimum for inelastic scattering) that can be imparted to a target nuclide $T$ by an incoming WIMP traveling with speed $v = \vmin$. In this case the recoil energy depends on the target nuclide. Here, we treat $\vmin$ as an independent variable, because this choice allows us to account for any isotopic target composition by summing terms dependent on $\ER^{T}(\vmin)$ over target nuclides $T$, for any fixed detected energy $\Ed$. 

Experiments do not actually measure the recoil energy of a target nucleus, but rather a proxy for the recoil energy that we call $\Ed$ (\eg the number of photoelectrons detected in a photomultiplier tube or some amount of ionization). The predicted measured differential rate as a function of the detected energy $\Ed$ involves a convolution of the differential recoil rate with the efficiency function $\epsilon(\ER,\Ed)$ and the energy resolution function $G_T(\ER,\Ed)$ of the experiment, which together give the probability that a detected recoil energy $\Ed$ resulted from a true recoil energy $\ER$, namely
\begin{equation}\label{eq:diffrate_ep}
\frac{\ud R}{\ud \Ed} = \sum_T \int_0^\infty \ud \ER \, \epsilon(\ER,\Ed) \, G_T(\ER,\Ed) \, \frac{\ud R_T}{\ud \ER} \, .
\end{equation}

We will now show that the predicted measured scattering rate can be expressed in terms of a convolution of two functions, one containing the astrophysical dependence and the other, which we call ``response function", containing the information on the DM particle physics and the detector response. Changing the order of integration in \Eq{eq:diffrate_ep}, the differential rate in detected energy can be written as
\begin{equation}\label{diffrate_manip1}
\frac{\ud R}{\ud \Ed} = \frac{\sigma_\text{ref} \rho}{m} \int \ud^3 v \, \frac{f(\bol{v},t)}{v} \, \frac{\ud \Hcur}{\ud \Ed} (\Ed, \bol{v}) \, ,
\end{equation}
where we used  a DM particle candidate and experiment dependent differential response function defined as
\begin{equation}\label{eq:dHcurlTotal'}
\frac{\ud \Hcur}{\ud \Ed} \equiv \sum_T  \frac{\ud \Hcur_T}{ \ud \Ed} \, , \nonumber
\end{equation}
\begin{equation}\label{eq:dHcurl}
\frac{\ud \Hcur_T}{\ud \Ed}(\Ed, \bol{v}) \equiv 
  \begin{dcases} 
      \hfill  \frac{C_T}{m_T}\int_{\ER^{T,-}}^{\ER^{T,+}} \, \ud \ER \epsilon(\ER,\Ed) \, G_T(\ER, \Ed) \, \frac{v^2}{\sigma_\text{ref}} \, \frac{\ud \sigma_T}{\ud \ER}(\ER, \bol{v})    \hfill & \text{ if $v \geqslant v_\delta^T$,} \\
      \hfill 0 \hfill & \text{ if $v < v_\delta^T$.} \\
  \end{dcases}
\end{equation}
Restricting ourselves to differential cross sections that only depend on the speed $v = |\bol{v}|$ of the incoming WIMP, the response function in~\Eq{eq:dHcurl} is only a function of the speed $v$. This occurs when the detector is isotropic and when the incoming WIMPs and target nuclei are unpolarized, as is most common. The parameter $\sigma_\text{ref}$  in \Eq{diffrate_manip1} and \Eq{eq:dHcurl}, is a factor extracted from the cross section to parametrize the overall strength of the interaction.  For example, in the differential cross section for the usual spin independent (SI) interaction,
\begin{equation}\label{eq:sigmaSI}
\frac{\ud \sigma_T^{SI}}{\ud \ER}(\ER,v) = \sigma_p \frac{\mu_T^2}{\mu_p^2}[Z_T+(A_T-Z_T)(f_n/f_p)]^2 \, \frac{F_T^2(\ER)}{2 \mu_T^2 v^2 / m_T} \, ,
\end{equation}
 the usual choice is $\sigma_\text{ref} = \sigma_p$, where  $\sigma_p$ is the WIMP-proton cross section.  In \Eq{eq:sigmaSI}  the factors $A_T$ and $Z_T$ are the atomic and charge numbers of nuclide $T$, $f_n$ and $f_p$ are the neutron and proton couplings, and $F_T(\ER)$ is the form factor normalized to $F_T(0)=1$ (in this case the Helm form factor).

Notice that writing the expected rate in \Eq{diffrate_manip1} the halo model  dependent quantities $\rho$ and $f({\bf v},t)$ are separated from the particle physics and
detector-dependent response function $\Hcur$. Using the speed distribution $F(v,t) \equiv v^2 \int \ud \Omega_v f(\bol{v},t)$, which is normalized to
1, \ie
\begin{equation}\label{eq:Fnormalization}
\int_0^\infty{\rm d}v~F(v,t)=1,
\end{equation}
the rate in \Eq{diffrate_manip1} can also be written as
\begin{equation}\label{drate_detatilde}
 \frac{\ud R}{\ud \Ed} = \frac{\sigma_{\rm ref}\rho}{m}\int_0^\infty{\rm d}v~\frac{F(v,t)}{v}\frac{{\rm d}\Hcur}{{\rm d}E'} (v) .
 \end{equation}
 Integrating over a particular detected energy range $(E'_1,E'_2)$ we obtain the
integrated rate
\begin{equation}\label{eq:integratedR}
  R_{[E'_1,E'_2]}=\frac{\sigma_{\rm ref}\rho}{m}\int_0^\infty{\rm d}v~\frac{F(v,t)}{v}\Hcur_{[E'_1,E'_2]}(v)
\end{equation}
where
\begin{equation}
  \Hcur_{[E'_1,E'_2]}(v)=\int_{E'_1}^{E'_2}{\rm d}E'\frac{{\rm d}\Hcur}{{\rm d}E'}(v).
\end{equation}

We can also use the halo function $\tilde{\eta}(\vmin,t)$ defined as
\begin{equation}\label{eq:eta_t}
\tilde{\eta}(\vmin,t) \equiv \frac{\rho \sigma_\text{ref}}{m}\int_{\vmin}^{\infty} \, \ud v \, \frac{F(v,t)}{v} \, ,
\end{equation}
to express the differential observed rate as
\begin{equation}
\frac{\ud R}{\ud \Ed} = \int_{0}^{\infty} \ud \vmin~ \tilde{\eta}(\vmin, t) \, \frac{\ud \mathcal{R}}{\ud \Ed}(\Ed, \vmin) \, ,
\end{equation}
using a different differential response function $\ud \mathcal{R}/ \ud \Ed$ which we introduce here to make contact with the formalism we used in previous papers, \eg\cite{Gondolo:2012rs, Gelmini:2015voa, Gelmini:2016pei}. This  DM particle model and detector dependent differential response function for $\tilde{\eta}(\vmin, t)$ is  related to  the previously defined differential response function for $f(\bol{v},t)/v$ defined in \Eq{eq:dHcurl} by 
\begin{equation}
\frac{\ud \mathcal{R}}{\ud \Ed}(\Ed,\vmin) \equiv \frac{\partial}{\partial \vmin}\left[ \frac{\ud \Hcur}{\ud \Ed}(\Ed,\vmin) \right] \, .
\end{equation}
Now the energy integrated rate in \Eq{eq:integratedR} can alternatively be written as
\begin{equation}
R_{[E'_1,E'_2]}=\int_0^\infty{\rm d}\vmin~ \tilde\eta(\vmin,t){\mathcal R}_{[E'_1,E'_2]}(\vmin) \, ,
\end{equation}
using the energy-integrated ${\mathcal R}_{[E'_1,E'_2]}(\vmin)$ response function for $\tilde{\eta}(\vmin, t)$
\begin{equation}
\mathcal{R}_{[\Ed_1,\Ed_2]}(\vmin) = \int_{\Ed_1}^{\Ed_2} \, \ud \Ed \, \frac{\ud \mathcal{R}}{\ud \Ed}(\Ed, \vmin) \, .
\end{equation}

We can now make a harmonic expansion of the halo function $\teta(\vmin,t)$, which is a function of time due to the  rotation of  Earth around the Sun
\begin{equation} \label{eta}
\tilde{\eta}(\vmin,t) \simeq \tilde{\eta}^0(\vmin) + \tilde{\eta}^1(\vmin) \cos(2 \pi (t-t_0)/\text{year}) + \dots  .
\end{equation}
Note that using \Eq{eq:eta_t} this is equivalent to an expansion of the speed distribution, \ie
\begin{equation}
F(v,t) \simeq F^0(v) + F^1(v) \cos(2 \pi (t-t_0)/\text{year}) + \dots \, ,
\end{equation}
which subsequently results in a harmonic expansion of the integrated rate, with expansion coefficients $R^{a}$ ($a= 0,1,2 \dots$) given by
\begin{equation}\label{eq:rateeqeta}
R^{a}_{[\Ed_1,\Ed_2]} = \int_{0}^{\infty} \ud \vmin \, \tilde{\eta}^a (\vmin) \, \mathcal{R}_{[\Ed_1,\Ed_2]}(\vmin) \, .
\end{equation}
or
\begin{equation}\label{eq:rateeq}
R^{a}_{[\Ed_1,\Ed_2]} = \frac{\sigma_{\rm ref}\rho}{m}\int_0^\infty{\rm d}v~\frac{F^{a}(v)}{v}\Hcur_{[E'_1,E'_2]}(v) \, .
\end{equation}
\Eq{eq:rateeq} will be used extensively in the discussion that follows.

The halo-independent anlaysis is carried out for a fixed dark matter particle model, including a fixed dark matter mass. This dependence enters into both the response functions and the relation between $\vmin$ and the recoil energy depend on the dark matter particle model (including the dark matter particle mass). 

In the following Secs.~\ref{Likelihood}  to \ref{sec:examples} 
we consider only unmodulated $a=0$ components of the rates, $F(v)$ and $\tilde\eta(v)$ functions although we drop the $a=0$ upper index everywhere for convenience 

\section{The Likelihood \label{Likelihood}}

We will generalize  here the formalism for the construction of halo-independent confidence bands using an extended likelihood function,  developed in \cite{Fox:2014kua, Gelmini:2015voa, Gelmini:2016pei}  to any likelihood function. Namely we will be able to work with any type of unmodulated rate datasets (\ie removing the necessity for unbinned data). 

Let us start by clarifying why the formalism of~\cite{Fox:2014kua, Gelmini:2015voa, Gelmini:2016pei} required an extended likelihood. The condition of having a non-increasing halo function was implemented using the Karush-Kuhn-Tucker (KKT) conditions. As shown in~\cite{Gelmini:2015voa}, one of these conditions implies that the halo function that maximized the likelihood must be constant except where the KKT multiplier is zero. With an extended likelihood, the KKT multiplier has only a finite number of isolated zeros~\cite{Gelmini:2015voa}, thus the halo function is piecewise constant with  a finite number of downward steps. The maximum number of isolated  zeroes of the KKT multiplier (and thus the number of downward steps of the halo function) is the number of observed events.  For a Poisson or Gaussian likelihood (which instead use binned data), one can easily show that the KKT multiplier can have extended zeroes, in which case the shape of the best-fit halo function is not determined by the KKT conditions. Without knowing the functional form of the best-fit halo function, one can only rely on purely numerical methods in which $\tilde{\eta}(\vmin)$ is approximated by a discretization with a very large number of bins to find the best fit function (see \eg~\cite{Feldstein:2014ufa}), however constructing the confidence band, which has not been done for binned, with this procedure quickly becomes numerically taxing. In \Sec{sec:theorems} we will prove that it is always possible to maximize any likelihood with a piecewise constant halo function with a maximum number of steps related to the total number of data entries, although this function may or may not be unique.

We begin by emphasizing that any likelihood depends on the halo model only through the predicted rates. For experiments with binned data, denoted with the index  $\alpha$,  $\alpha=1,\dots,n_{\rm binned}$, one usually uses 
either a Poisson likelihood
\begin{equation}
{\cal L}_\alpha[\tilde\eta]=
\prod_{j=1}^{N_{bin-\alpha}}
\frac{(\nu_{\alpha j}[\tilde\eta]+b_{\alpha j})^{n_{\alpha j}}}{n_{\alpha j}!}
\exp[-(\nu_{\alpha j}[\tilde\eta]+b_{\alpha j})],
\end{equation}
or a Gaussian likelihood
\begin{equation}\label{eq:Lgauss}
{\cal L}_\alpha[\tilde\eta]=
\prod_{j=1}^{N_{bin-\alpha}}
\frac{1}{\sigma_{\alpha j}\sqrt{2\pi}}
\exp[-(\nu_{\alpha j}[\tilde\eta]+b_{\alpha j}-n_{\alpha j})^2/\sigma_{\alpha j}^2],
\end{equation}
where $\nu_{\alpha j}[\tilde\eta]$ is the number of events predicted in the bin $j$ of experiment $\alpha$, i.e.
\begin{equation}
\nu_{\alpha j}[\tilde\eta]=(MT)_\alpha R_{\alpha j}[\tilde\eta] \, ,
\end{equation}
$R_{\alpha j}$ is the integrated predicted rate in the same bin, and $(MT)_\alpha$ is the exposure of the experiment $\alpha$.

For experiments with unbinned data, denoted here with the index $\beta$,  $
\beta=1,2,\dots,n_{\rm unbinned}$,  one can use an extended likelihood,
\begin{equation}
{\cal L}_\beta[\tilde\eta]=e^{-\nu_\beta[\tilde\eta]}\prod_{j=1}^{N_{O\beta}}(MT)_\beta
\left.\left(\frac{{\rm d}R_\beta}{{\rm d}E'}[\tilde\eta]+\frac{{\rm d}R_{\beta b}}{{\rm d}E'}\right)\right|_{E'=E'_j},
\end{equation}
where ${\rm d}R_\beta/{\rm d}E'$ is the predicted differential rate, ${\rm d}R_{\beta b}/{\rm d}E'$ is the background differential rate, $N_{O\beta}$ is the total number of observed events, and $\nu_\beta$ is the total number of expected events, \ie
\begin{equation}\label{eq:nutotal}
\nu_\beta[\tilde\eta]=(MT)_\beta(R_\beta[\tilde\eta])_{\rm total} \, ,
\end{equation}
which depends on the exposure $(MT)_\beta$ of the experiment $\beta$ and the total energy integrated rate $(R_\beta)_{\rm total}$ predicted for experiment $\beta$.

It is clear from the previous equations of this section that the likelihoods  depend on the halo model only through the differential or integrated expected rates. 

We will prove in the next section that all predicted unmodulated rates can be written in terms of a speed distribution $F(v)$ consisting of a sum of a finite number of delta functions in speed $v$. Thus we  will  always be able to maximize any likelihood using this functional form for $F(v)$.

For convenience, let us define a different functional of $\tilde\eta$ (and thus a functional also of $F(v)$), the $-2$ log-likelihood functional
\begin{equation}\label{eq:Ldefinition}
L[\tilde\eta]=-2\ln{\cal L}[\tilde\eta].
\end{equation}
With this definition, maximizing the likelihood is equivalent to minimizing $L$.

\section{${F(v)}$ given as a sum of delta functions in speed \label{sec:theorems}}

We will work in a vector space of dimension ${\mathcal N}$ in which each vector $\vec R$ has as components a complete set of possible predicted rates $\{R_k\}$; these could be: \emph{(i)} energy integrated rates in bins $i=1,2,\dots,N_{bin-\alpha}$ of experiments $\alpha=1,2,\dots,n_{\rm binned}$, \ie
\begin{equation}\label{eq:Rbinned}
R_k\equiv R_{\alpha i}=\left(\frac{\rho\sigma_{\rm ref}}{m}\right)\int_0^\infty{\rm d}v~\Hcur_{\alpha i}(v)\frac{F(v)}{v},
\end{equation}
\emph{(ii)} a differential rate for the event with observed energy $E'_j$, $j=1,2,\dots,N_{O\beta}$ of experiments $\beta=1,2,\dots,n_{\rm unbinned}$, \ie
\begin{eqnarray}\label{eq:Runbinned}
R_k\equiv R_{\alpha i}&=&\Delta E'\left.\frac{{\rm d}R_\beta}{{\rm d}E'}\right|_{E'=E'_j}\nonumber\\
&=&\Delta E'\left(\frac{\rho\sigma_{\rm ref}}{m}\right)\int_0^\infty{\rm d}v~\left.\frac{{\rm d}\Hcur_{\alpha i}(v)}{{\rm d}E'}\right|_{E'=E'_j}\frac{F(v)}{v},
\end{eqnarray}
or \emph{(iii)} the total predicted rate over the whole energy interval of an experiment with unbinned data, \ie
\begin{equation}\label{eq:Rtotal}
R_k\equiv (R_{\beta})_{\rm total}=\left(\frac{\rho\sigma_{\rm ref}}{m}\right)\int_0^\infty{\rm d}v~\Hcur_{\beta}(v)\frac{F(v)}{v}.
\end{equation}
Here we multiply the differential rate by a fixed energy, e.g. $\Delta E'=1$ keV, so that all components $R_k$ have the same dimension and can thus be summed (as we will do later in this section).

Therefore, each rate vector $\vec R=(R_1,R_2,\dots,R_{\mathcal N})$ has ${\mathcal N}$ components, where ${\mathcal N}$ is total number of rate data points
\begin{equation}\label{eq:curlyN}
  {\mathcal N} =\sum_{\alpha=1}^{n_{\rm binned}} N_{{\rm bin}-\alpha}
  + \sum_{\beta=1}^{n_{\rm unbinned}} (N_{O\beta}+1).
\end{equation}
The factor $(\rho\sigma_{\rm ref}/m)={\mathcal C}$ is a constant factor common to all components of a rate vector. Notice that ${\mathcal C}F(v)$ incorporates all the rate dependence on the halo model. Let us define also the vector $\vec\Hcur$ whose components are either (see \Eq{eq:Rbinned})
\begin{equation}\label{eq:Hvec1}
\Hcur_k(v)=\Hcur_{\alpha i}(v)
\end{equation}
(see \Eq{eq:Rbinned}), or 
\begin{equation}\label{eq:Hvec2}
\Hcur_k(v)=\Delta E'\left.\frac{{\rm d}\Hcur_{\alpha i}(v)}{{\rm d}E'}\right|_{E'=E'_j}
\end{equation}
(see \Eq{eq:Runbinned}), or
\begin{equation}\label{eq:Hvec3}
\Hcur_k(v)=\Hcur_\beta(v) 
\end{equation}
(see \Eq{eq:Rtotal}).
Thus we can write
\begin{equation}\label{eq:Rvector}
\vec R= {\mathcal C}\int_0^\infty{\rm d}v~\frac{\vec\Hcur(v)}{v}F(v).
\end{equation}

Let us now define a convex hull of the rate vectors. Within the vector space of rate vectors $\vec R$ we can define the {\it convex hull} of any finite set of $K$ vectors, which are called {\it generating vectors} (i.e. those vectors that generate the hull) denoted here with a superscript $(j)$, $\vec R^{(j)}$, $j=1,2,\dots,K$. The {\it convex hull} of this set of generating vectors is the set of all vectors $\vec R$ which are a {\it convex combination} of the generating vectors, \ie a linear combination with non-negative real coefficients $\lambda_j$ which sum to $1$. In other words, the convex hull contains all vectors $\vec R= \sum_{j=1}^K\lambda_j \vec R^{(j)}$ with $\lambda_j$ real and $\lambda_j\geq0$, and with $\sum_{j=1}^K\lambda_j=1$. The dimension $d$ of the convex hull is $d\leq{\mathcal N}$. For example, the hull of 3-dimensional generating vectors contained on a plane is a surface, i.e. has dimension $d=2$ (see \Fig{fig:caratheodory_discrete} for examples).

Instead of choosing a discrete set of $K$ vectors $\vec R^{(j)}$ to define the hull, we start with a continuous line of vectors ${\mathcal C}\vec\Hcur(v)/v$ with a continuous non-negative label $v$ (the speed). All vectors $\vec R$ in the hull are given in \Eq{eq:Rvector} as a convex combination, which is an integration in this case, i.e. a linear combination of ${\mathcal C}\vec\Hcur(v)/v$ with real non-negative coefficients $F(v)$ normalized to 1 ($\int_0^\infty F(v){\rm d}v=1$, as it corresponds to the speed distribution). 
Passing from a discrete to a continuous set of generating vectors, and thus using an integral representation of the vectors in a convex hull, as in \Eq{eq:Rvector}, is justified by the Choquet theorem \cite{choquet1956existence}.

By identifying the rate vectors as elements of a convex hull, we can now use well-established mathematical theorems.
The Caratheodory theorem (see Appendix A.1) says that any vector $\vec R$ in the convex hull of dimension $d$ of a generating set of vectors belongs to the convex hull of at most $(d+1)$ of the generating vectors, \ie any $\vec R$ can be written as the convex combination of at most $(d+1)$ generating vectors.

The Caratheodory theorem applies to the convex hull of any set of generating vectors.
However, when the generating vectors are a connected set, as is our case, the Fenchel-Eggleston theorem (see Appendix A.2) reduces the maximum number of required generating vectors. Thus, any vector $\vec R$ defined by \Eq{eq:Rvector} can be written as a linear combination of only at most $d$ of the generating vectors, namely ${\mathcal C}\vec\Hcur(v_h)/v_h$ with $h=1,2,\dots,d$ where $d$ is the dimension of the convex hull. Thus
\begin{equation}\label{eq:Rfinite}
\vec R=\sum_{h=1}^{d}\frac{{\mathcal C}\vec\Hcur(v_h)}{v_h}F_h
\end{equation}
with real non-negative coefficients $F_h$ such that
\begin{equation}\label{eq:Fhsum1}
\sum_{h=1}^dF_h=1.
\end{equation}
Notice that this is equivalent to saying that for a particular DM particle candidate and detector, any rate can be written in terms of a speed distribution consisting of a sum of at most $d$ delta functions in speed
\begin{equation}\label{eq:Fdeltas}
F(v)=\sum_{h=1}^dF_h\delta(v-v_h).
\end{equation}
Notice that the normalization condition \Eq{eq:Fnormalization} on the speed distribution $F(v)$ coincides with the condition given in \Eq{eq:Fhsum1}.

So far we have worked with rate vectors $\vec R$ of dimension ${\mathcal N}$, and thus the dimension of the convex hull we define is $d\leq {\mathcal N}$. However, because all physically meaningful rates are non-negative, we can use the fact that the sum of all vector components $R_k$ of any rate vector $\vec R$ is positive, and define the vectors
\begin{equation}
\hat R=\frac{\vec R}{\sum_{k=1}^{\mathcal N}R_k},
\end{equation}
which are the projections of the rate vectors $\vec R$ on to the plane
\begin{equation}
\sum_{k=1}^{\mathcal N}\hat R_k=1
\end{equation}
with dimension $({\mathcal N}-1)$. As shown in the appendix, using the $\hat R$ vectors the results shown above still hold, i.e. \Eq{eq:Rfinite} and \Eq{eq:Fdeltas} are still valid, except now $d\leq{\mathcal N}-1$.

Thus for the purpose of finding the rates $\vec R$ that maximize the likelihood we can always take the speed distribution $F(v)$ as a sum of at most $d={\mathcal N}-1$ delta function in speed \Eq{eq:Fdeltas}. If the likelihood includes at least one extended likelihood this result coincides with previous findings in \cite{Gelmini:2016pei} that we expressed in terms of $\tilde\eta$,
\begin{equation}
\tilde\eta(\vmin)={\mathcal C}\int_{\vmin}^\infty\frac{F(v)}{v}{\rm d}v.
\end{equation}

We find again that using \Eq{eq:Fdeltas} $\tilde\eta(\vmin)$ is a piecewise constant non-increasing functions of $\vmin$ with at most $d\leq {\mathcal N}-1$ downward steps. It is important to note the value of $d$, \ie the maximum possible number of downward steps of the best-fit halo function, cannot be known \emph{a priori}, it is only possible to bound this number from above by ${\mathcal N}-1$. The actual number of downward steps must be determined on a case-by-case basis by numerically minimizing the log-likelihood starting with a piecewise constant halo function with ${\mathcal N}-1$ downward steps.
This maximum step-counting agrees with our previous results of \cite{Gelmini:2015voa, Gelmini:2016pei}, where it was found that when the likelihood contains at least one extended likelihood the best-fit $\tilde\eta_{BF}(\vmin)$ function is of this form with the same number of maximum steps (and we further proved that it is unique \cite{Gelmini:2016pei}), although our previous counting \cite{Gelmini:2016pei} applied to likelihoods with exactly one extended likelihood.

An equation similar to~\Eq{eq:Fdeltas} is also found in a recent paper using linear programming techniques for halo-independent comparisons of direct and indirect DM searches~\cite{Ibarra:2017mzt}. The Eq.(14) of~\cite{Ibarra:2017mzt} is similar to our~\Eq{eq:Fdeltas}. Although it is written in terms of delta functions in velocity, it is clear that when treating rate (not modulation) data it is the speed distribution that is taken in~\cite{Ibarra:2017mzt}  to be a sum of delta functions in speed, as in ~\Eq{eq:Fdeltas}. The maximum number of terms in the summation in~\cite{Ibarra:2017mzt}  may appear to be different at a first glance, because it is given in terms of the number of experiments (instead of data entries as in our equation), but in~\cite{Ibarra:2017mzt} only one bin is used per experiment, thus the number of experiments and the number of bins coincide. It seems clear that if more than one bin would be used per experiment the maximum number of terms in Eq.(14) would become the total number of bins, but it is much less clear to us how the formalism in~\cite{Ibarra:2017mzt}  could be extended to consider unbinned data (i.e. how to write Eqs.  11 or 12 of ~\cite{Ibarra:2017mzt} for each single event). However, the major difference in the approach of ~\cite{Ibarra:2017mzt} and ours, is that the functions to be minimized or maximized with constraints in~\cite{Ibarra:2017mzt}  are rates, and not the likelihood as we do. In our language of convex hulls, ~\cite{Ibarra:2017mzt} looks for restricted regions in the hull of all possible rates and examines their boundaries. The purpose of both approaches is also clearly different: in~\cite{Ibarra:2017mzt} the purpose if to obtain limits on the  parameters defining the DM particle model. The formalism of~\cite{Ibarra:2017mzt} does not attempt to produce regions of halo models compatible with data, as we do. Thus, in spite of the connections of linear programming with the geometrical properties of convex hulls, the intent and application of the methods presented in~\cite{Ibarra:2017mzt} and those of this work to analyze time-averaged rates differ significantly. Another clear difference is the way in which the method of~\cite{Ibarra:2017mzt} is applied to modulation amplitude data, which will be the subject of our~\Sec{sec:modulation}. In this case it is the velocity distribution in the Galactic rest frame that is written in terms of a sum of delta functions (\ie a sum of `streams' with negligible velocity dispersion). The method in~\cite{Ibarra:2017mzt} is applied to account for the modulation data of DAMA, combined with other data sets. In~\cite{Ibarra:2017mzt} the annual modulation amplitude of each stream is defined as half the difference between the rate at December $1^{\rm st}$ and June $1^{\rm st}$, however this is not the correct definition of the annual modulation amplitude as it implicitly assumes a sinusoidal modulation with a particular phase (namely the modulation that would be found from the Standard Halo Model). Streams, however, very rarely produce such a modulation. The authors of~\cite{Ibarra:2017mzt} are aware of this problem but  do not address  how to correctly analyze measurements of the rate modulation within such an analysis. In this respect, the approach we introduce in~\Sec{sec:modulation}  is very different.

\section{How to find $\tilde\eta_{BF}$ and determine if it is unique} \label{sec:bf}

We have proven that any likelihood can be maximized (or, equivalently,  the corresponding $L$ functional defined in \Eq{eq:Ldefinition} can be minimized) using
 a piecewise constant $\tilde\eta$ function with at most $({\mathcal N}-1)$ downward steps. Thus, we can define a function $f_L^{({\mathcal N}-1)}$ of $2({\mathcal N}-1)$ variables specifying the positions and heights of the ${\mathcal N}-1$ steps, $\vec{v}=(v_1,v_2,\dots,v_{{\mathcal N}-1})$ and $\vec{\tilde\eta}=(\tilde\eta_1,\tilde\eta_2,\dots,\tilde\eta_{{\mathcal N}-1})$, as a restriction of the functional $L[\tilde\eta]$, namely replacing  $\tilde\eta$ by a piecewise constant function in the likelihood:
\begin{equation}
f_L^{({\mathcal N}-1)}(\vec{v},\vec{\tilde\eta})
\equiv
L[\tilde\eta^{({\mathcal N}-1)}(\vmin;\vec{v},\vec{\tilde\eta})].
\label{eq:fL_NO}
\end{equation}
The piecewise constant function $\tilde\eta^{({\mathcal N}-1)}$ is defined as
\begin{eqnarray}
\tilde\eta^{({\mathcal N}-1)}(\vmin;\vec{v},\vec{\tilde\eta})
\equiv
  \begin{cases}
    \tilde\eta_a&\text{if}~ v_{a-1} \leq \vmin < v_a,\\
    0&\text{if}~ v_{{\mathcal N}-1} \leq \vmin,
  \end{cases}
 \label{eq:etaNO}
\end{eqnarray}
where $a=1, \dots, {\mathcal N}-1$ and we define $v_0=0$ (note that \Eq{eq:etaNO} requires the definition of $v_0$). Here we assume $\vmin$ and  all the $v_a$'s are larger than $v_\delta$ (defined here as the smallest of all $v_\delta^T$ scattering thresholds  for all  nuclides $T$ in all experiments  included in the likelihood)
 and the constraints $\tilde\eta_a\leq\tilde\eta_b$ for $a>b$ are satisfied.
Since the function $\tilde\eta$ cannot change after the last step and it must reach zero for large $\vmin$, it must be zero for $\vmin>v_{{\mathcal N}-1}$.
We do not specify the value of $\tilde\eta^{({\mathcal N}-1)}$ below the minimum $v_\delta$ since the event rate is independent of it.

From these definitions, we can easily obtain $\tilde\eta_{\rm BF}$ and $L_{\rm min}$, the minimum value of the functional $L[\tilde\eta]$, by finding the positions and hights of the steps $\vec v_{\rm BF}$ and $\vec{\tilde\eta}_{\rm BF}$ that minimize $f_L^{({\mathcal N}-1)}$, so that\
\begin{equation}
\tilde\eta_{\rm BF}(\vmin)=\tilde\eta^{({\mathcal N}-1)}(\vmin;\vec v_{\rm BF},\vec{\tilde\eta}_{\rm BF})
\end{equation}
and
\begin{eqnarray}
L_{\rm min}&\equiv&L[\tilde\eta_{\rm BF}(\vmin)]=L[\tilde\eta^{({\mathcal N}-1)}(\vmin;\vec v_{\rm BF},\vec{\tilde\eta}_{\rm BF})].
\end{eqnarray}

The minimization of the function $f_L^{({\mathcal N}-1)}$ of $2({\mathcal N}-1)$ parameters $v_1,\dots,v_{{\mathcal N}-1},\:\tilde\eta_1,\dots,\tilde\eta_{{\mathcal N}-1}$, subject to the constraints 
\begin{eqnarray}
 &v_1&> v_\delta, 
 \label{eq:constr_v_positive} \\
 &v_b&-v_a \ge 0 \text{ and }\: \tilde\eta_a - \tilde\eta_b \ge 0 \:\text{ for } a<b,
 \label{eq:constr_steps}
\end{eqnarray}
can be done numerically using a global minimization algorithm.
In the implementation, we express $f_L^{({\mathcal N}-1)}$ in terms of $\ln \tilde\eta_a$, and use $\ln \tilde\eta_a$ instead of $\tilde\eta_a$ as variables, since $\tilde\eta_a$ span many orders of magnitude.
This also accounts for the $\tilde\eta_a \geq 0$ constraints, leaving only the constraints in \Eq{eq:constr_v_positive} and \Eq{eq:constr_steps} to be enforced in the minimization.

The authors have previously proven, in Appendix B of \cite{Gelmini:2016pei} {\footnote{Notice that the proof in Appendix A.3 of~\cite{Gelmini:2016pei} is not correct, because it does not take into account higher order derivatives of the likelihood. However this was a redundant proof with another in Appendix A.1, which is correct.}}, that if the likelihood contains at least one extended likelihood the $\tilde\eta_{BF}$ function is unique.  This proof relies on analyzing the behavior of the function
\begin{equation}\label{eq:qmin1}
q(\vmin)=\int_0^\infty{\rm d}v\frac{\delta (-2\ln{\cal L})}{\delta\tilde\eta(v)} \,
 = \sum_\alpha q_{\alpha}(\vmin) ,
\end{equation}
defined in Eq.~(3.7) and (3.8) of \cite{Gelmini:2016pei} as
\begin{equation}\label{Qpoisson}
q_{\alpha}(\vmin) \equiv 2 \sum \limits_{j=1}^{N_{bin-\alpha}} \left[\frac{\nu_{\alpha j}[\tilde{\eta}] + b_{\alpha j} - n_{\alpha j}}{\nu_{\alpha j}[\tilde{\eta}] +b_{\alpha j}}\right] (MT)_\alpha\Hcur_{\alpha j}(\vmin) \, 
\end{equation}
for an experiment $\alpha$ using a Poisson likelihood, and    
\begin{equation}\label{Qgauss}
q_{\alpha}(\vmin) \equiv 
2 \sum \limits_{j=1}^{N_{bin-\alpha}}
\left[ \frac{\nu_{\alpha j}[\tilde{\eta}] + b_{\alpha j} - n_j^{\alpha j}}{\sigma_j^2} \right]
(MT)_\alpha\Hcur_{\alpha j}(\vmin) \, ,
\end{equation}
for an experiment $\alpha$ using  a Gaussian likelihood. Specifically, if one can show that this function $q(\vmin)$ in \Eq{eq:qmin1} is not zero over an extended $\vmin$ interval, then the proof of \cite{Gelmini:2016pei} still applies and the best-fit halo function is unique. Furthermore, it is argued in \cite{Gelmini:2016pei} that solutions with exact cancellations between different bins are unphysical, and since $\Hcur_{\alpha j}$ are strictly positive functions above some $\vmin$ value $v^*_{\alpha j}$, obtaining an extended zero over some interval $[v_1, v_2]$ requires $\nu_{\alpha j}= n_{\alpha j} -b_{\alpha j}$ for all $\alpha j$ with $v_{\alpha j}^* < v_2$. Note that $v_{\alpha j}^*$ roughly corresponds to the smallest value of $\vmin$ to which a particular bin $\alpha j$ has sensitivity. If the above condition is not satisfied, \ie $\nu_{\alpha j} \neq n_{\alpha j} -b_{\alpha j}$ for all $\alpha j$ with $v_{\alpha j}^* < v_2$, the proofs in \cite{Gelmini:2016pei} guarantee that $\tilde\eta_{BF}$ is unique. We describe in the next section how one can in practice determine if the aforementioned conditions are met.

We will show that for a likelihood exclusively comprised of Poisson or Gaussian likelihoods the $\tilde\eta_{BF}$ function is not guaranteed to be unique.

\section{Confidence and degeneracy bands}\label{sec:cband}

In \cite{Gelmini:2015voa, Gelmini:2016pei}, we defined two-sided pointwise confidence band with the following procedure. 

The halo-independent confidence band can be defined as the region filled by all possible $\teta$ functions satisfying
\begin{equation}\label{CBdef}
\Delta L[\teta] \equiv L[\teta] - L_\text{min} \leq \Delta L^* \, ,
\end{equation}
where $L_\text{min}$ is the minimum of $L[\teta]$, and $\Delta L^*$ corresponds to a desired confidence level (we show below that the relevant probability distribution approaches the chi-square distribution with one degree of freedom in the limit of large data samples). In practice, finding all $\teta$ functions satisfying \Eq{CBdef} is not possible. Instead, let us consider the subset of $\teta$ functions which minimize $L[\teta]$ subject to the constraint of passing by a particular point $(v^*,\tilde\eta^*)$,
\begin{equation}\label{constrain}
\teta(v^*) = \teta^* \, .
\end{equation}
Now let us define $L_\text{min}^{c}(v^*,\teta^*)$ to be the minimum of $L[\teta]$ subject to the constraint in \Eq{constrain}, and
\begin{equation}\label{eq:DeltaLcmin}
\Delta L_\text{min}^{c}(v^*,\teta^*) = L_\text{min}^{c}(v^*,\teta^*) - L_\text{min} = -2\ln\left[\frac{\underset{\tilde\eta(v^*)=\tilde\eta^*}{\max}{\cal L}[\tilde\eta]}{\max{\cal L[\tilde\eta]}}\right]\, ,
\end{equation}
where in the square bracket is the ratio of the constrained maximum value of the likelihood and the unconstrained maximum value of the likelihood, which is always smaller or equal to 1. If the point $(v^*,\teta^*)$ lies within the confidence band, then there should exist at least one $\teta$ function passing through this point which satisfies $\Delta L[\teta] \leq \Delta L^*$. Should this be the case, it follows that $\Delta L_\text{min}^{c}(v^*,\teta^*) \leq \Delta L^*$. Alternatively, if $\Delta L_\text{min}^{c}(v^*,\teta^*) \geq \Delta L^*$, one can state that there does not exist a single $\teta$ which satisfies $\Delta L[\teta] \leq \Delta L^*$. Thus the confidence band can be constructed by finding the values of $(v^*,\teta^*)$ which satisfy $\Delta L_\text{min}^{c}(v^*,\teta^*) \leq \Delta L^*$. This condition defines a two-sided interval around $\tilde\eta_{\rm BF}$ for each $\vmin$ value (with $\vmin=v^*$), and the collection of those intervals forms a pointwise confidence band in $\vmin$--$\tilde\eta$ space, which we  simply call the confidence band. 

{One can also use this procedure to identify what we refer to as a {\it degeneracy band} by taking $\Delta L^*$ to zero. This band contains degenerate best-fit halo functions and there are no such functions outside the band. Namely, if $\Delta L_\text{min}^{c}(v^*,\teta^*) > 0$, one can state that there does not exist a single $\teta$ passing through  the $(v^*,\teta^*)$ point that satisfies $\Delta L[\teta] = 0$.

We will prove now that the $\teta$ functions which minimize $L[\teta]$ subject to the constraint in \Eq{constrain}, have at most ${\mathcal N}+1$ steps instead of at most ${\mathcal N}$ steps. Let us  implement the constraint by introducing a fictitious extra component $R_{{\mathcal N}+1}$ of the rate vectors $\vec R$ in \Eq{eq:Rvector}, so that the dimensionality of the rate vector space becomes ${\mathcal N}+1$ instead of 
${\mathcal N}$.   
 Using the unit of rate $({\rm kg}\cdot{\rm day})^{-1}$ and ${\rm day}^{-1}$ as the unit of $\tilde\eta$, the additional component $\Hcur_{{\mathcal N}+1}$ of the $\vec\Hcur(v)$ vector is defined by
\begin{equation}
\Hcur_{{\mathcal N}+1}(v)\equiv ({\rm kg})^{-1}\Theta(v-v^*)\, .
\end{equation}

Thus using \Eq{eq:Rvector}, we get
\begin{equation}\label{eq:RNp1}
R_{{\mathcal N}+1}={\mathcal C}\int_0^\infty{\rm d}v~\frac{F(v)}{v}\Hcur_{{\mathcal N}+1}(v)
={\mathcal C}({\rm kg})^{-1}\int_{v^*}^\infty{\rm d}v~\frac{F(v)}{v}
=({\rm kg})^{-1}\tilde\eta(v^*)\, .
\end{equation}
Now the constraint in \Eq{constrain} becomes 
\begin{equation}
R_{{\mathcal N}+1}=({\rm kg})^{-1}\tilde\eta^*\, .
\end{equation}
Notice that now any rate vector in the convex hull of ${\mathcal C}\vec\Hcur(v)/v$ can be rewritten as in \Eq{eq:Rfinite}, as a sum of at most ${\mathcal N}$ terms (instead of ${\mathcal N}-1$). Therefore, we can find the constrained minimum $L^c_{\rm min}(v^*,\tilde\eta^*)$ with the same procedure we specify in \Sec{sec:bf} to find unconstrained minimum $L_{\rm min}$, but with the maximum number of steps in $\tilde\eta$ increased by one. Notice that, differently from what we found in \cite{Gelmini:2015voa} and \cite{Gelmini:2016pei}, the potential extra step of $\tilde\eta$ is not necessarily located at $\vmin=v^*$. We will address this point shortly.

 If the $\tilde\eta_{BF}$ is unique, \ie if there is no degeneracy band, Wilks theorem can be applied as in \cite{Gelmini:2015voa} and \cite{Gelmini:2016pei}. In order to understand the meaning of $\Delta L^c_{\rm min}$, we can discretize the continuous variable $\vmin$ into a collection of $K$ discrete values
 $\vec{v}_\text{min} = (\vmin^0,\dots,\vmin^{K-1})$. The functional $L[\tilde\eta]$ in \Eq{eq:Ldefinition} then becomes a function of the $K-$dimensional vector $\vec{\teta} = (\teta_0,\teta_1,\dots,\teta_{K-1})$ which defines the piecewise constant function $\teta(\vmin;\vec{\teta})$ given by
\begin{equation}
\teta(\vmin;\vec{\teta}) \equiv \teta_i \text{  if  } \vmin^i \leq \vmin < \vmin^{i+1} \, .
\end{equation}
With this discretization, the constraint on $(v^*,\teta^*)$ in \Eq{constrain} corresponds to $\vmin^k \leq v^* < \vmin^{k+1}$ and $\teta^* = \teta_k$ for some integer $0 \leq k \leq K-1$. $\Delta L_\text{min}^{c}(v^*,\teta^*)$ is then replaced by the function $\Delta L_\text{min}^{k,c}(\teta^*)$ with the index $k$ corresponding to $v^*$, defined by
\begin{equation}\label{consproflike}
\Delta L_\text{min}^{k,c}(\teta^*) = - 2 \ln \left[\frac{\mathcal{L}(\hat{\hat{\teta}}_0,\dots,\hat{\hat{\teta}}_{k-1},\teta_k=\teta^*,\hat{\hat{\teta}}_{k+1},\dots,\hat{\hat{\teta}}_{K-1})}{\mathcal{L}(\hat{\teta}_0,\dots,\hat{\teta}_k,\dots,\hat{\teta}_{K-1})} \right] \, , 
\end{equation}
where $\hat{\hat{\teta}}_i$ are the $\teta_i$ values which maximize the likelihood function 
\begin{equation}
\mathcal{L}(\teta_0,\dots,\teta_{K-1}) \equiv \mathcal{L}[\teta(\vmin; \vec{\teta})]
\end{equation}
 subject to the constraint $\teta_k = \teta^*$, and $\hat{\teta}_i$ maximize $\mathcal{L}$ without the constraint. $\Delta L_\text{min}^{k,c}(\teta^*)$ now defines the $-2\ln$ of the profile likelihood ratio with one parameter ($\teta_k$), and thus by Wilks' theorem the distribution of $\Delta L_\text{min}^{k,c}(\teta^*)$ approaches the chi-square distribution with one degree of freedom in the limit where the data sample is very large. If we now recover the continuum limit by taking $K \rightarrow \infty$, we see that $\Delta L_\text{min}^{k,c}(\teta^*)$ approaches $\Delta L_\text{min}^{c}(v^*,\teta^*)$. Thus the construction of the confidence band is equivalent to finding the collection of confidence intervals in $\teta^*$ for each $v^*$ at a given CL for which $\Delta L^c_\text{min} < \Delta L^*$. Assuming that $\Delta L_\text{min}^c$ is chi-square distributed, the choices $\Delta L^* = 1.0$ and $\Delta L^* = 2.7$ correspond to the confidence intervals of $\teta$ at $68\%$ and $90\%$ CL, respectively, for each $\vmin$ value (see \Fig{fig:HI_XeI}).

If the vertical width of the degeneracy band is non-zero, there are multiple $\tilde\eta$ functions which give the same maximum value of the likelihood. In this case Wilks theorem does not apply, thus we do not know the probability distribution of $\Delta L^c_{\rm min}$. In Fig.~\ref{fig:HI_XeD} we still show the bands corresponding to $\Delta L^*=1.0$ and $\Delta L^*=2.7$, but they do not correspond to a particular CL.

\section{Examples of data analysis}\label{sec:examples}

In this section, we are going to present examples of the methods developed above to mock data. While it is preferable to demonstrate these methods using real experimental results, currently the only binned experiment with putative signal is DAMA/LIBRA.
However, DAMA/LIBRA observed the modulation while the methods presented above apply to unmodulated rates.
For this reason, we illustrate these techniques using two mock xenon-based experiments that have been crafted so as to demonstrate both the `unique best-fit' and `not-unique best-fit' cases presented above. We hereby refer to these mock experiments as `Xe-D' and `Xe-I', where the `D' and `I' are used to differentiate whether the number of observed events decreases or increases as bin energies increase.

Xe-D is a xenon-based experiment with a one ton-year exposure. It contains three adjacent bins spanning observed event energies $[0.5, 1.5]$ keV, $[1.5, 3]$ keV, and $[3, 4.5]$ keV. These bins are assumed to have observed 6 events, 4 events, and 1 event (in increasing order of $\ER$), while the expected background for each bin is 1 event. This experiment is assumed to have a Gaussian energy resolution with $\sigma = 0.15$ keV\footnote{This number is taken to be large enough that information on the derivative of $\teta(\vmin)$ can be used in the numerical minimization procedure and small enough to not introduce excessive overlap between bins.}. Xe-D is assumed to have perfect detection efficiency at all energies. The binning and data described for Xe-D are depicted in the form of a histogram in \Fig{fig:fake_data} (shown in pink). 

Xe-I is also a xenon-based experiment with a one ton-year exposure. It contains three adjacent bins spanning observed event energies $[1, 2.5]$ keV, $[2.5, 4]$ keV, and $[4, 5.5]$ keV. These bins are assumed to have observed 1 event, 4 events, and 6 events (in increasing order of $\ER$), while the expected background for each bin is again 1 event. As with Xe-D, perfect efficiency is assumed at all recoil energies, and the energy resolution is taken to be a Gaussian with $\sigma = 0.15$ keV. The binning and data described for Xe-I are also depicted in \Fig{fig:fake_data} (shown in gray). 

\begin{figure}
\centering
\includegraphics[width=.49\textwidth]{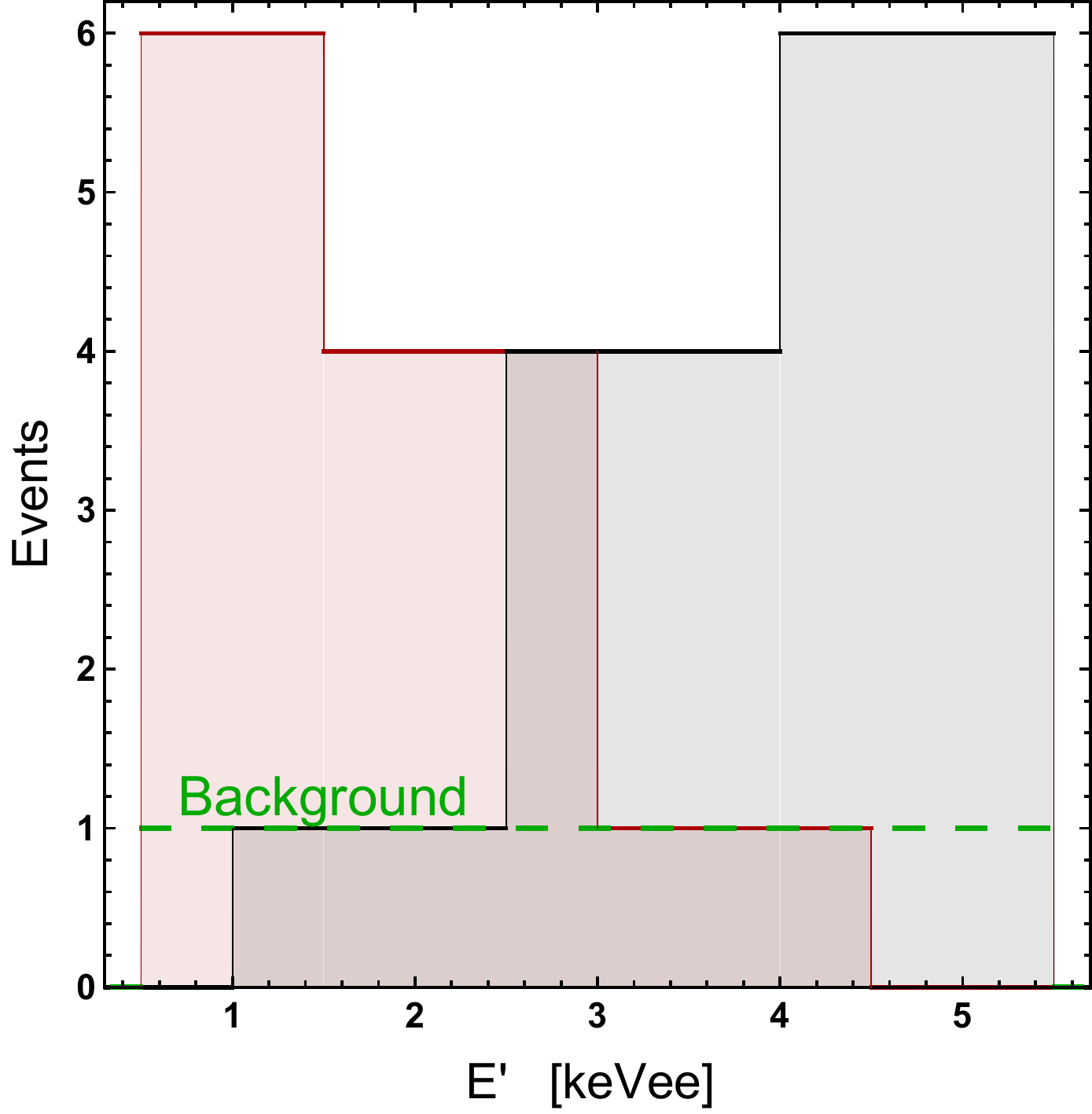}
\caption{\label{fig:fake_data} Binning scheme and mock data used in the analysis presented in \Sec{sec:examples}. Three bins are used for both `Xe-D' (shown in pink) and 'Xe-I' (shown in grey). The predicted background number of background events for both experiments and each bin is taken to be one (depicted with horizontal dashed green line). The number of events observed, in order of increasing energy bins, is 6 (1), 4 (4), an 1 (6) for Xe-D (Xe-I). }
\end{figure}

Assuming a  piecewise constant best-fit halo function with at most ${\mathcal N}-1=2$ steps, we maximize a Poisson likelihood assuming a 9 GeV DM particle and an elastic isospin conserving spin-independent interaction. The determined best-fit piecewise constant halo is depicted with a thick green line in \Fig{fig:HI_XeD}. As expected, this best-fit halo function produces 5, 3, and 0 events in first, second and third bin, respectively. The predicted number of events are exactly the measured minus the background in each bin, thus the function $q_{\alpha}(\vmin)$ in \Eq{Qpoisson} is zero, the arguments of Sect. \ref{sec:bf} lead us to
 suspect that the best-fit is not unique. Thus we proceed by calculating at each $(v^*, \teta^*)$ the constrained best-fit halo function. As expected, there exist large regions of parameter space for which the value of the constrained likelihood equals the value of the best-fit likelihood. In \Fig{fig:HI_XeD}, we identify the region (green band) of parameter space for which the maximum constrained likelihood is within $10^{-3}$ of the maximum likelihood. In principle this value of $\Delta L$ should be 0, however numerical errors that arise from the minimization procedure limit the precision of the calculation of $\Delta L$ to approximately this level. Finally, we use the previously calculated constrained likelihood to identify confidence bands corresponding to $\Delta L = 1.0$ (darker yellow) and $\Delta L = 2.7$ (lighter yellow). Had the best-fit halo function been unique, Wilks theorem would have applied and one would expect these confidence bands to approach the $68\%$ and $90\%$ CL, respectively.
Notice that in the figures we lose information below the lowest energy bin, and the confidence band is progressively unbounded from above. In \Fig{fig:HI_XeD}, this happens below $\vmin=150$ km/s. Since the purpose of plotting these bands is usually to compare the compatibility of putative and null signals, we extend $\teta_{BF}$ in our plots to this region, in the most conservative way (\ie constant) although $\tilde\eta_{BF}$ is actually undetermined. The same happens in \Fig{fig:HI_XeI} and \Fig{fig:HI_XeCombo} below $250$ km/s and $150$ km/s, respectively.

\begin{figure}
\centering
\includegraphics[width=.49\textwidth]{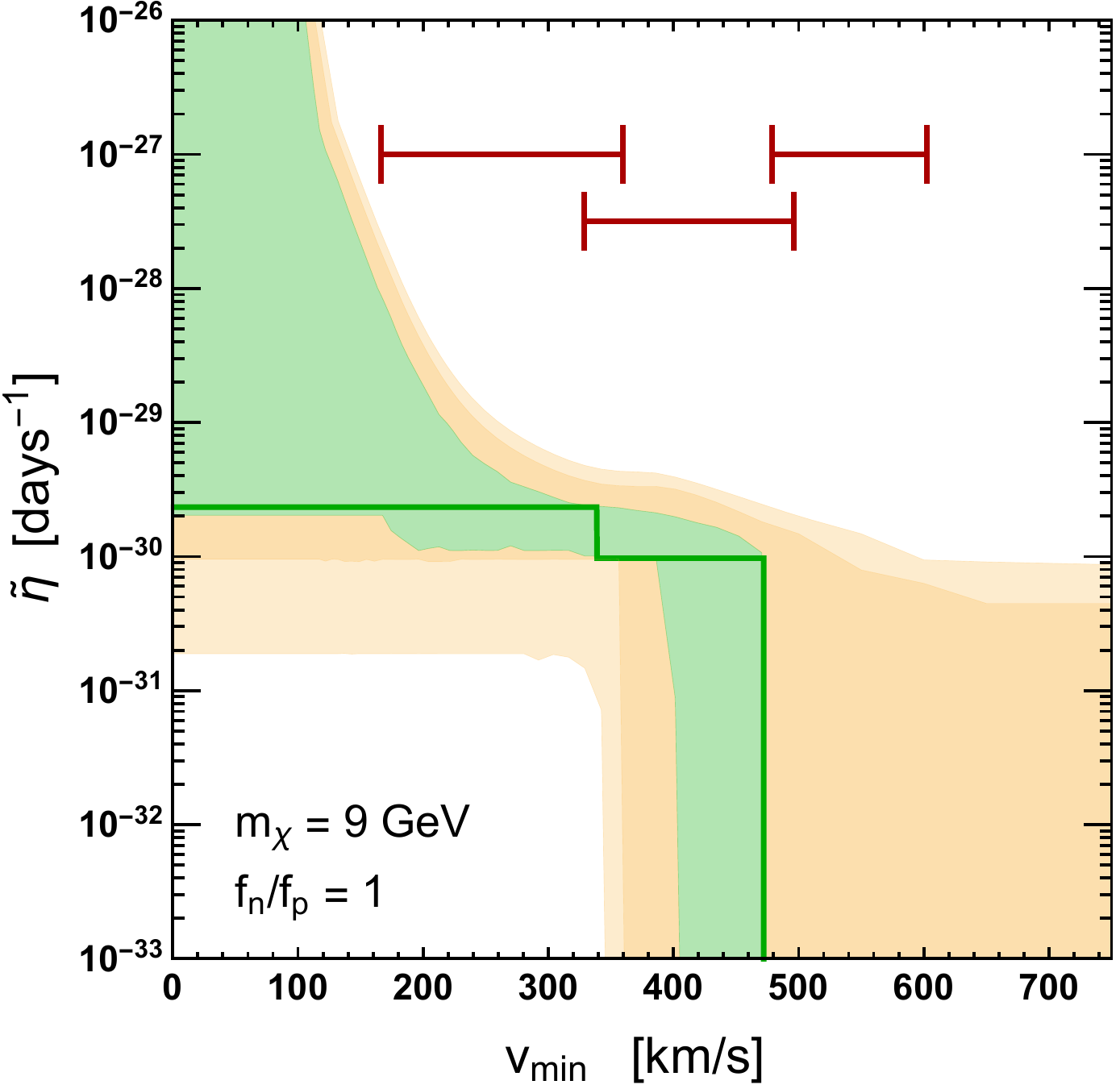}
\caption{\label{fig:HI_XeD}
 Halo-independent analysis for Xe-D, assuming an elastic spin-independent contact interaction with isospin conserving couplings, and a 9 GeV DM particle. The best-fit halo function derived using a piecewise constant halo function is shown with a thick green line; this best-fit halo function for Xe-D is determined not to be unique, and thus the degeneracy band (green region) is found using the technique outlined in \Sec{sec:examples}. Two-sided pointwise confidence bands are depicted for $\Delta L = 1.0$ (darker yellow) and $\Delta L = 2.7$ (lighter yellow), which however do not correspond to defined CL. The horizontal red lines indicate the $\vmin$ ranges where each bin's response function $\mathcal{R} \equiv \partial_{v_{\rm min}}\mathcal{H}$ is significantly non-zero are. For a detected energy bin $[E_1, E_2]$ they show the range $[\min \left(\vmin(E_1 - \sigma) \right), \max \left(\vmin(E_2 + \sigma) \right)]$ where $\min$ and $\max$ refer to the minimum and maximum $\vmin$ values for the respective recoil energy among all xenon isotopes. }
\end{figure}

In \Fig{fig:HI_XeI}, the aforementioned procedure is applied to the Xe-I data. Here, the global maximum of the likelihood, assuming each bin is described by a Poisson likelihood, occurs when 0, 3, and 5 DM events are predicted in the first, second, and third bin, respectively. It is clearly not possible to realize this global maximum of the likelihood using a monotonically decreasing halo function (assuming again an elastic SI contact interaction), and thus the function $q_{\alpha}(\vmin)$ in \Eq{Qpoisson} does not have extended zeroes, and may have at most isolated zeroes (where the steps of the $\teta_{BF}$ functions will be located). In this case the arguments of Sect. \ref{sec:bf} tell us that the best-fit halo function is unique. Indeed, maximizing this likelihood using a piecewise constant halo function with 2 steps, and assuming once again a 9 GeV elastic spin-independent interaction, confirms that this best-fit halo function is unique. This result is illustrated in \Fig{fig:HI_XeI}. Since the best-fit halo function is unique, the methods of Refs.~\cite{Gelmini:2015voa,Gelmini:2016pei} can be applied, and a two-sided pointwise confidence band can be constructed. The $68\%$CL (darker blue) and $90\%$CL (lighter blue) confidence bands are shown for this model in \Fig{fig:HI_XeI}.

\begin{figure}
\centering
\includegraphics[width=.49\textwidth]{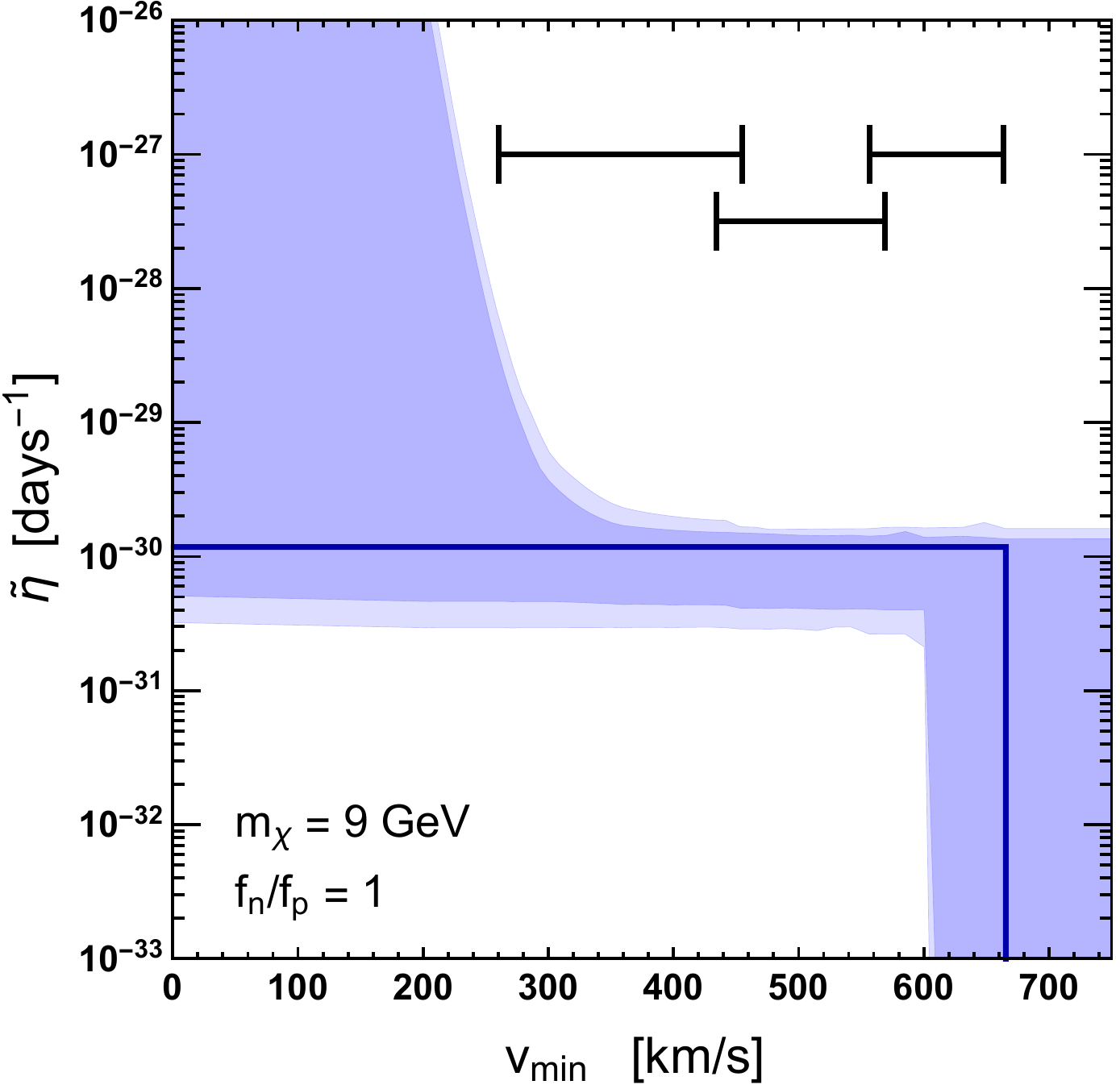}
\caption{\label{fig:HI_XeI} Halo-independent analysis for Xe-I, assuming an elastic spin-independent contact interaction with isospin conserving couplings, and a 9 GeV DM particle. The best-fit halo function derived using a piecewise constant halo function is shown with a thick blue line; this best-fit halo function for Xe-I is determined to be unique, and the $68\%$CL (darker blue) and $90\%$CL (lighter blue) two-sided pointwise confidence bands  are shown (see~\Sec{sec:examples}). The horizontal black lines indicate the $\vmin$ ranges where each bin's response function $\mathcal{R} \equiv \partial_{v_{\rm min}}\mathcal{H}$ is significantly non-zero. For a detected energy bin $[E_1, E_2]$ they show the range $[\min \left(\vmin(E_1 - \sigma) \right), \max \left(\vmin(E_2 + \sigma) \right)]$ where $\min$ and $\max$ refer to the minimum and maximum $\vmin$ values for the respective recoil energy among all xenon isotopes. }
\end{figure}

It should be clear that as the total number of bins in the likelihood increases, the relative ease with which a monotonically decreasing halo function can simultaneously maximize all individual binned likelihoods decreases; this additional strain on the likelihood increases the chance of obtaining a unique best-fit halo function. To illustrate this point, we consider in \Fig{fig:HI_XeCombo} a joint likelihood analysis of Xe-D and Xe-I, again assuming an elastic spin-independent interaction and a 9 GeV DM particle. Here, the best-fit halo function is found to be unique. As before, the constrained likelihood is then used to construct $68\%$ and $90\%$ CL confidence bands.

\begin{figure}
\centering
\includegraphics[width=.49\textwidth]{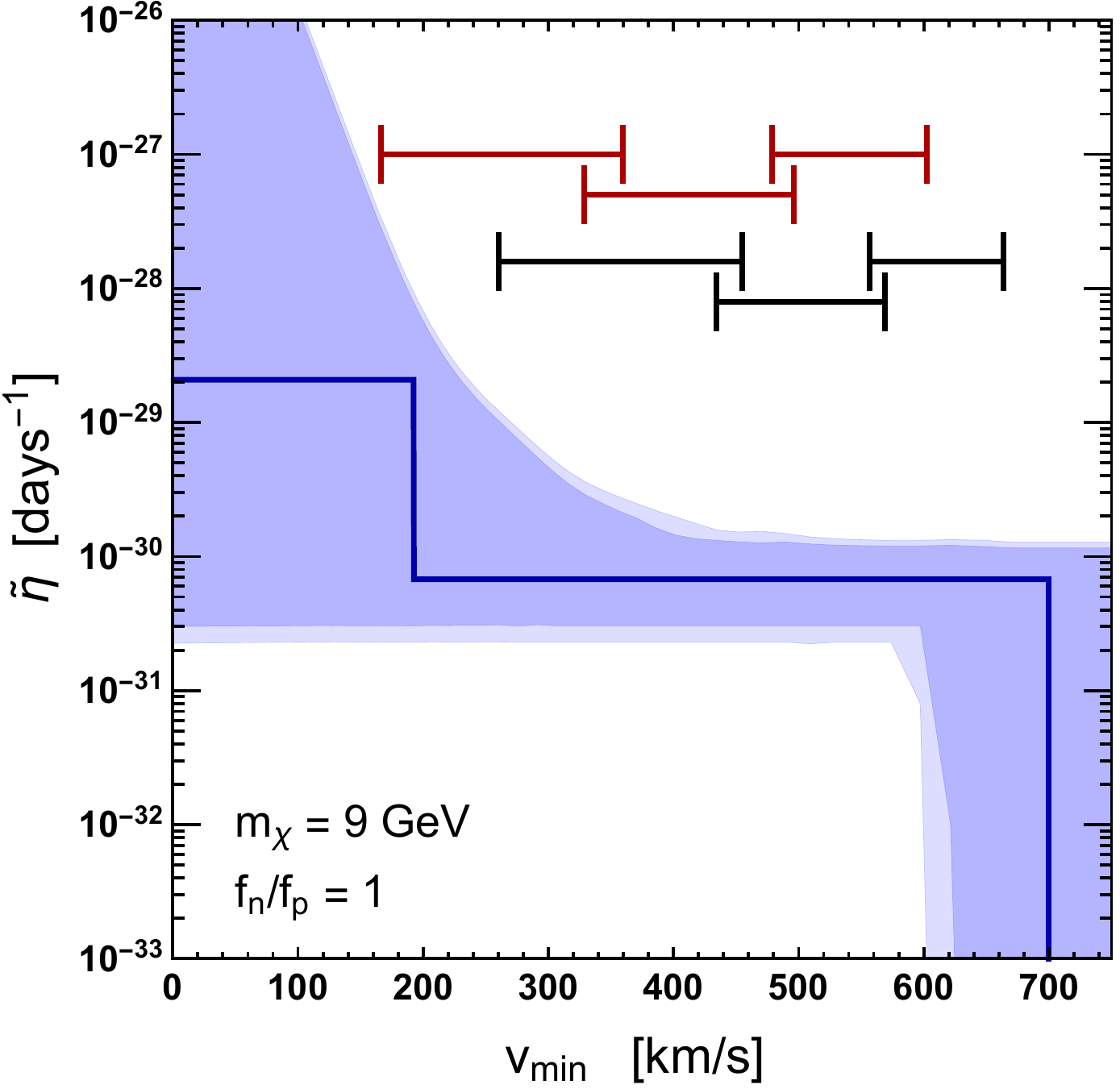}
\caption{\label{fig:HI_XeCombo} Same as \Fig{fig:HI_XeI}, but for the joint analysis of Xe-D and Xe-I. The best-fit halo function in this case is found to be unique.   }
\end{figure}

\section{Halo-independent analysis of modulation amplitudes}\label{sec:modulation}

In this section, we show how to extend the formalism presented so far, based on the use of the Fenchel-Eggleston theorem, to the analysis of all the coefficients of the harmonic expansion of the time-dependent rate.
The main idea is to shift the time dependence of the velocity distribution function to the detector response functions, by making a change of reference frame from Earth's frame to the Galactic frame.
This can always be done, but it can be complicated when gravitational focusing of DM by the Sun (see e.g. \cite{Alenazi:2006wu}), or other Earth position-dependent effects, are important. We are going to neglect these effects, and assume here that the change of frame can be done with a Galilean velocity transformation. 
Then, using the normalization to $1$ of the velocity distributions, we can express the rates as vectors in the convex hull of time-dependent response function vectors.

The time-dependent rate $R_{\alpha i}(t)$ in bin $i$ of experiment $\alpha$ can be expressed as
\begin{equation}\label{eq:R3}
R_{\alpha i}(t)=\int{\rm d}^3v~{\mathscr H}_{\alpha i}(\vec v)f(\vec v,t)\, .
\end{equation}
Here the definition of the ${\mathscr H}_{\alpha i}(\vec v)$ function differs from our $\Hcur$ function by a factor of ${\mathcal C}/v$, \ie
\begin{equation}
{\mathscr H}_{\alpha i}(\vec v)=\frac{{\mathcal C}\Hcur_{\alpha i}(\vec v)}{v}\, .
\end{equation}
(recall that ${\mathcal C}\equiv\rho\sigma_{\rm ref}/m$).
If we were to take the time-average of $R_{\alpha i}$ and $f(\vec v,t)$ in \Eq{eq:R3}, we would recover \Eq{eq:Rbinned} and our previous analysis.
Here, instead we will assume a Galactic DM velocity distribution $f^{\rm gal}(\vec u)$.
Ref.~\cite{Gondolo:2017jro} shows that $R_{\alpha i}(t)$ can also be written as
\begin{equation}\label{eq:Rmodulation}
R_{\alpha i}(t)=\int{\rm d}^3u~{\mathscr H}^{\rm gal}_{\alpha i}(\vec u,t)~f^{\rm gal}(\vec u)\, ,
\end{equation}
where $\vec u=\vec v_\odot+\vec v_\oplus(t)+\vec v$ is the WIMP velocity with respect to the Galaxy, $\vec v$ is the usual WIMP velocity with respect to the Earth, $\vec v_\odot$ is the velocity of the Sun with respect to the Galaxy and $\vec v_\oplus(t)$ is Earth's velocity with respect to the Sun. The laboratory velocity distribution $f(\vec v,t)$ is related to the Galactic velocity distribution by
\begin{equation}
f(\vec v,t)=f^{\rm gal}(\vec v_\odot + \vec v_\oplus(t) + \vec v)\, .
\end{equation}
Notice that here $f^{\rm gal}(\vec u)$ does not depend on time, so the time dependence of the rate $R_{\alpha i}(t)$ comes from the experiment and particle candidate dependent response function ${\mathscr H}_{\alpha i}^{\rm gal}(\vec u,t)$.
\Eq{eq:Rmodulation} makes it clear that the time dependence of the rate is due to the periodic change in the motion of the detector in the Galaxy.

Now we can generate the (halo-dependent) convex hull of the connected set of the vectors $\overrightarrow{\mathscr H}^{\rm gal}(\vec u,t)$, defined as in \Eq{eq:Hvec1} to \Eq{eq:Hvec3} but using the vector $\vec u$ as the continuous label instead of $v$.
The rate vectors $\vec R(t)$ in the hull are defined as in \Eq{eq:Rbinned} to \Eq{eq:Rtotal} but using \Eq{eq:Rmodulation}, namely
\begin{equation}\label{eq:Rt}
\vec R(t)=\int{\rm d}^3u~\overrightarrow{\mathscr H}^{\rm gal}(\vec u,t)~f^{\rm gal}(\vec u)\, ,
\end{equation}
where $f^{\rm gal}(\vec u)$ is normalized to $1$, \ie
\begin{equation}
\int{\rm d}^3u~f^{\rm gal}(\vec u)=1\, .
\end{equation}

Writing $\vec R(t)$ and $\overrightarrow{\mathscr H}^{\rm gal}(\vec u,t)$ as a harmonic series in time
\begin{equation}\label{eq:Rt_harmonic}
\vec R(t)=\vec R^0+\vec R^1\cos(\omega(t-t_0))+\cdots\, ,
\end{equation}
and
\begin{equation}
\overrightarrow{\mathscr H}^{\rm gal}(\vec u,t)~=~\overrightarrow{\mathscr H}^{{\rm gal}-0}(\vec u)~+~\overrightarrow{\mathscr H}^{{\rm gal}-1}(\vec u)~\cos(\omega(t-t_0))~+~\cdots\, ,
\end{equation}
we can express \Eq{eq:Rt} in terms of the expansion coefficients for $a=0,1,\dots$,
\begin{equation}\label{eq:Ravec}
\vec R^a=\int{\rm d}^3u~\overrightarrow{\mathscr H}^{{\rm gal}-a}(\vec u)~f^{\rm gal}(\vec u)\, .
\end{equation}
The vectors $\vec R^a$ form the convex hull of the connected set of generating vectors $\overrightarrow{\mathscr H}^{{\rm gal}-a}(\vec u)$.
We can now apply the Fenchel-Eggleston theorem to this hull, so any vector $\vec R^a$ can be written as
\begin{equation}\label{eq:Rasum}
\vec R^a=\sum_{h=1}^d\overrightarrow{\mathscr H}^{{\rm gal}-a}(\vec u_h)~f_h^{{\rm gal}-a}\, ,
\end{equation}
with the non-negative real coefficients $f_h^{{\rm gal}-a}$ satisfying $\sum_{h=1}^d f_h^{{\rm gal}-a}=1$.
Here $d$ is the dimension of the hull.
Equivalently, the Galactic velocity distribution can be written as a sum of at most $d$ delta functions in Galactic velocity $\vec u$, namely streams with zero velocity dispersion,
\begin{equation}\label{eq:fgalsum}
f^{\rm gal}(\vec u)=\sum_{h=1}^d f_h^{{\rm gal}-a}~\delta^{(3)}(\vec u-\vec u_h^a)\, .
\end{equation}
Depending on the symmetries of $\overrightarrow{\mathscr H}^{{\rm gal}-a}(\vec u)$ as function of $\vec u$ the dimension of the connected generating set could be $3$ or less (it is parameterized by a $3$-dimensional vector $\vec u$). In any event, the dimension $d$ of the hull is $d\leq{\mathcal N}$, where ${\mathcal N}$ is the number of components of the vectors $\overrightarrow{\mathscr H}^{\rm gal}(\vec u,t)$. For the $\overrightarrow{\mathscr H}^{{\rm gal}-a}$ vectors whose components are not positive definite, we cannot use the projection explained in Appendix B, which reduces the maximum necessary number of terms from $d$ to $d-1$.
This reduction can only be done for $a=0$.
\Eq{eq:Ravec}, \Eq{eq:Rasum}, and \Eq{eq:fgalsum} for $a=1$ correspond to Eq.~(3.10), (3.13) and (3.12) of \cite{Gondolo:2017jro} except that the maximum number of terms in the latter equations is ${\mathcal N}+1$ (our ${\mathcal N}$ is the same as $N$ in \cite{Gondolo:2017jro}).
In our case we could replace $d$ by ${\mathcal N}$ assuming that some of the $f_h^{{\rm gal}-a}$ coefficients are zero.

In order to find the best-fit velocity distribution $f^{\rm gal}_{BF}(\vec u)$,  we proceed as in \Sec{sec:bf}, but using the expression in \Eq{eq:fgalsum}.
We find the set of $4{\mathcal N}$ parameters, $\vec u^a_1,\vec u^a_2,\dots,\vec u^a_{{\mathcal N}}$ and $f^{{\rm gal}-a}_1,f^{{\rm gal}-a}_2,\dots,f^{{\rm gal}-a}_{\mathcal N}$, that minimizes the function $f_L$ defined by using \Eq{eq:fgalsum} in the likelihood functional, namely
\begin{equation}
f_L\left(\vec u^a_1,\dots,\vec u^a_{{\mathcal N}};
f^{{\rm gal}-a}_1,\dots,f^{{\rm gal}-a}_{{\mathcal N}}\right)
\equiv \left.L\left[f^{\rm gal}\right]\right|_{f^{\rm gal}=\sum_{h=1}^{{\mathcal N}}f_h^{{\rm gal}-a}\delta^{(3)}(\vec u-\vec u_h^a)}\, .
\end{equation}
For $a\neq0$, we define the Gaussian likelihood as in \Eq{eq:Lgauss} but for modulation amplitudes.
Notice that the likelihood functional ${\cal L}[\tilde\eta]$ is also a functional ${\cal L}[f^{\rm gal}]$ of the Galactic velocity distribution $f^{\rm gal}(\vec u)$, and thus so is the functional $L=-2\ln{\mathcal L}$.

With the function $f^{\rm gal}_{BF}(\vec u)$ in the form of set of streams as in \Eq{eq:fgalsum}, the resulting $\teta_{BF}(\vmin,t)$ function
\begin{eqnarray}\label{eq:etaBF_t}
\teta_{BF}(\vmin,t)
&\equiv&
{\mathcal C}~\int_{|\vec v|\geq \vmin}{\rm d}^3v~\frac{f^{\rm gal}(\vec v_\odot+\vec v_\oplus(t)+\vec v)}{v}\nonumber\\
&=&
\sum_{h=1}^{{\mathcal N}}
\frac{{\mathcal C}~f^{{\rm gal}-a}_h}{|\vec u_h^a-\vec v_\odot-\vec v_\oplus(t)|}\Theta(|\vec u_h^a-\vec v_\odot-\vec v_\oplus(t)|-\vmin)\, ,
\end{eqnarray}
for any fixed time $t$ is a piecewise constant function with at most ${\mathcal N}$ downward steps (see dashed blue lines in the left panel of \Fig{fig:etas}).
We could now proceed to define a piecewise degeneracy or confidence band at any given time.
However, we prefer to use the time-average $\teta^0_{BF}(\vmin)$
\begin{equation}\label{eq:etaBF_0}
\tilde\eta^0_{BF}(\vmin)\equiv \frac{1}{T}\int{\rm d}t~\tilde\eta_{BF}(\vmin,t)
={\mathcal C}~\sum_{h=1}^{{\mathcal N}}\frac{f^{{\rm gal}-a}_h}{\overline{v_h}(\vmin)}
\, ,
\end{equation}
where  $\overline{v_h}(\vmin)$ is defined by
\begin{eqnarray}
\frac{1}{\overline{v_h}(\vmin)}&\equiv&\frac{1}{T}\int{\rm d}t~\frac{\Theta(|\vec u_h^a-\vec v_\odot-\vec v_\oplus(t)|-\vmin)}{|\vec u_h^1-\vec v_\odot-\vec v_\oplus(t)|}\, ,
\end{eqnarray}
with the time period $T$ of $1$ year, and construct the bands around it to make contact with our previous results in which we use only time-averaged rate data.
An example of $\teta^0_{BF}$ is shown by the black line in \Fig{fig:etas}.a.

To construct the confidence band and degeneracy band from the modulation amplitude data, it is necessary to impose the constraint \Eq{constrain}, \ie $\tilde\eta^0(v^*)=\tilde\eta^*$, in the minimization of the functional $L$.
As in \Sec{sec:cband}, this can be implemented by introducing an additional fictitious component $R_{{\mathcal N}+1}$ to the rate vector, but now we write it as an integral over the Galactic $3$-dimensional velocity $\vec u$ instead of the speed $v$ in the Earth's frame as done in \Eq{eq:RNp1}, as
\begin{equation}
R_{{\mathcal N}+1} \equiv \int{\rm d}^3u~\overrightarrow{\mathscr H}^{{\rm gal}-0}(\vec u)~f^{\rm gal}(\vec u)
=({\rm kg})^{-1}\tilde\eta(v^*)\, ,
\end{equation}
where
\begin{equation}
\overrightarrow{\mathscr H}^{{\rm gal}-0}(\vec u)
\equiv\frac{{\mathcal C}\cdot{\rm kg}^{-1}}{T}\int{\rm d}t~\frac{\Theta(|\vec u-\vec v_\odot-\vec v_\oplus(t)|-v^*)}{|\vec u-\vec v_\odot-\vec v_\oplus(t)|}\, .
\end{equation}
Introducing this fictitious rate simply allows for the correct counting of the maximum number of delta functions in \Eq{eq:fgalsum}, which becomes $({\mathcal N}+1)$.

Therefore, the constrained minimum $L^c_{\rm min}(v^*,\tilde\eta^*)$ of the functional $L[f^{\rm gal}]$ can be obtained by minimizing the function $f_L$ of $4({\mathcal N}+1)$ parameters,
\begin{equation}
f_L\left(\{\vec u^a_1,\dots,\vec u^a_{{\mathcal N}+1}\};\{f^{{\rm gal}-a}_1,\dots,f^{{\rm gal}-a}_{{\mathcal N}+1}\}\right)
\equiv \left.L\left[f^{\rm gal}\right]\right|_{f^{\rm gal}=\sum_{h=1}^{{\mathcal N}+1}f_h^{{\rm gal}-a}\delta^{(3)}(\vec u-\vec u_h^a)}\, ,
\end{equation}
subject to the constraint
\begin{equation}
\tilde\eta^* = {\mathcal C}\sum_{h=1}^{{\mathcal N}+1}f_h^{{\rm gal}-a}\frac{1}{T}\int{\rm d}t~\frac{\Theta(|\vec u_h^a-\vec v_\odot-\vec v_\oplus(t)|-v^*)}{|\vec u_h^a-\vec v_\odot-\vec v_\oplus(t)|}\, .
\end{equation}

With the constrained minimum $L^c_{\rm min}(v^*,\teta^*)$, we can define $\Delta L^c_{\rm min}(v^*,\teta^*)$ as in \Eq{eq:DeltaLcmin}, and find a region in $\vmin-\teta$ space satisfying $\Delta L(v^*,\tilde\eta^*)\leq\Delta L^*$, which we identify as either a degeneracy band by setting $\Delta L^*=0$, or a confidence band for a certain CL, for which we can find the corresponding $\Delta L^*$ value if the best-fit $\tilde\eta_{BF}$ function is unique.

Since the procedure just outlined using up to $4({\mathcal N}+1)$ parameters can be numerically challenging, we could consider simplifying the problem by restricting the type of velocity distributions included in the analysis.
If the velocity distribution is isotropic in the Galactic frame, namely if it only depends on the speed $u=|\vec u|$,
\begin{equation}
f^{\rm gal}(\vec u)=f^{\rm gal}(u)\, ,
\end{equation}
\Ref{Gondolo:2017jro} finds (see Eq.~A.2 of \cite{Gondolo:2017jro}) that time-dependent rates can be written in terms of just $u$,
\begin{equation}
R_{\alpha i}(t)=\int_0^\infty{\rm d}u~\overline{\mathscr H}^{\rm gal}_{\alpha i}(u,t)~F^{\rm gal}(u)\, ,
\end{equation}
where $\overline{\mathscr H}^{\rm gal}_{\alpha i}(u,t)$ is the angle averaged Galactic response function
\begin{equation}
\overline{\mathscr H}^{\rm gal}_{\alpha i}(u,t)\equiv \frac{1}{4\pi}\int{\rm d}\Omega_u~{\mathscr H}_{\alpha i}^{\rm gal}(\vec u,t)\, ,
\end{equation}
and $F^{\rm gal}(u)$ is the Galactic speed distribution, $F^{\rm gal}(u)=4\pi u^2f^{\rm gal}(u)$. In this case the generating connected set of vector $\overrightarrow{\overline{\mathscr H}}^{\rm gal}(\vec u,t)$ depends again, as in \Sec{sec:theorems}, on one continuous real parameter, which is here $u$ instead of $v$. Then all the time-dependent rate vectors
\begin{equation}\label{eq:Rvecgal}
\vec R(t)=\int_0^\infty{\rm d}u~\overrightarrow{\overline{\mathscr H}}^{\rm gal}(u,t)~F^{\rm gal}(u)\, ,
\end{equation}
constitute the convex hull of $\overrightarrow{\overline{\mathscr H}}^{\rm gal}(u,t)$, since the non-negative coefficients $F^{\rm gal}(u)$ sum to $1$, due to the normalization of the speed distribution, \ie $\int_0^\infty{\rm d}u~F^{\rm gal}(u)=1$.

Using a harmonic expansion of the time-dependent rate vector $\vec R(t)$ as in
 \Eq{eq:Rt_harmonic}, and the vector $\overrightarrow{\overline{\mathscr H}^{\rm gal}}(u,t)$,
\begin{equation}
\overrightarrow{\overline{\mathscr H}}^{\rm gal}(u,t)~=~\overrightarrow{\overline{\mathscr H}}^{{\rm gal}-0}(u)~+~\overrightarrow{\overline{\mathscr H}}^{{\rm gal}-1}(u)~\cos(\omega(t-t_0))~+~\cdots\, ,
\end{equation}
we can express \Eq{eq:Rvecgal} in terms of the expansion coefficients for $a=0,1,\dots$,
\begin{equation}\label{eq:Ravec_iso}
\vec R^a=\int{\rm d}u~\overrightarrow{\overline{\mathscr H}}^{{\rm gal}-a}(u)~F^{\rm gal}(u)\, .
\end{equation}
The vectors $\vec R^a$ form the convex hull of the connected set of generating vectors $\overrightarrow{\overline{\mathscr H}}^{{\rm gal}-a}(u)$.
We can now apply the Fenchel-Eggleston theorem to this hull, so any vector $\vec R^a$ can be written as
\begin{equation}\label{eq:Rasum_iso}
\vec R^a=\sum_{h=1}^{\mathcal N}\overrightarrow{\overline{\mathscr H}}^{{\rm gal}-a}(u_h)~F_h^{{\rm gal}-a}\, ,
\end{equation}
with the non-negative real coefficients $f_h^{{\rm gal}-a}$ satisfying $\sum_{h=1}^d f_h^{{\rm gal}-a}=1$. Equivalently, the Galactic speed and velocity distribution can be written as a sum of at most $d$ delta functions in Galactic speed $u$,
\begin{equation}\label{eq:Fgalsum}
F^{\rm gal}(u)=\sum_{h=1}^{\mathcal N} F_h^{{\rm gal}-a}~\delta(u-u_h^a)\, ,
\end{equation}
and
\begin{equation}
f^{\rm gal}(\vec u) = \sum_{h=1}^{\mathcal N}\frac{F^{{\rm gal}-a}_h}{4\pi u_h^2}\delta(u-u^a_h).
\end{equation}
Therefore, the time-dependent best-fit halo function $\tilde\eta_{BF}(\vmin,t)$ is now
\begin{eqnarray}
\tilde\eta_{BF}(\vmin,t)&\equiv&\int_{|\vec v|\geq\vmin}{\rm d}^3v~
{\mathcal C}~\frac{f^{\rm gal}(\vec v_\odot+\vec v_\oplus(t)+\vec v)}{|\vec v|}\nonumber\\
&=&{\mathcal C}~\sum_{h=1}^{\mathcal N}\frac{F_h}{4\pi u_h^2}~\int_{|\vec v|\geq\vmin}{\rm d}^3v~\frac{\delta^{(3)}(\vec v_\odot+\vec v_\oplus(t)+\vec v)}{|\vec v|}\, .
\end{eqnarray}
Thus
\begin{equation}\label{eq:etaBF_iso}
\tilde\eta_{BF}(\vmin,t)=\sum_{h=1}^{\mathcal N}~{\mathcal C}~F_h
\times
\begin{dcases} 
\frac{1}{u_h}\, &\text{$\vmin\leq u_h-u_\oplus(t)$}\\
\frac{u_\oplus(t)+u_h-\vmin}{2u_\oplus(t)u_h}\, &\text{$u_h-u_\oplus(t)<\vmin<u_h+u_\oplus(t)$}\\
0\, &\text{$u_h+u_\oplus(t)\leq\vmin$}\, ,
  \end{dcases}
\end{equation}
with $u_\oplus(t)$ defined as the time-dependent speed of the Earth in the Galactic frame $u_\oplus(t)=|\vec v_\odot+\vec v_\oplus(t)|$.
This is a piecewise non-increasing linear function, an example of which, using a single delta function at $u_1=300$ km/s, can be found in \Fig{fig:etas}.b.
We could with this expression continue our analysis by defining an average $\teta^0_{BF}$ function (an example of which is also found in \Fig{fig:etas}.b) and constructing either a degeneracy or confidence band using $2({\mathcal N}+1)$ parameters.

\begin{figure}
\centering
\includegraphics[width=.49\textwidth]{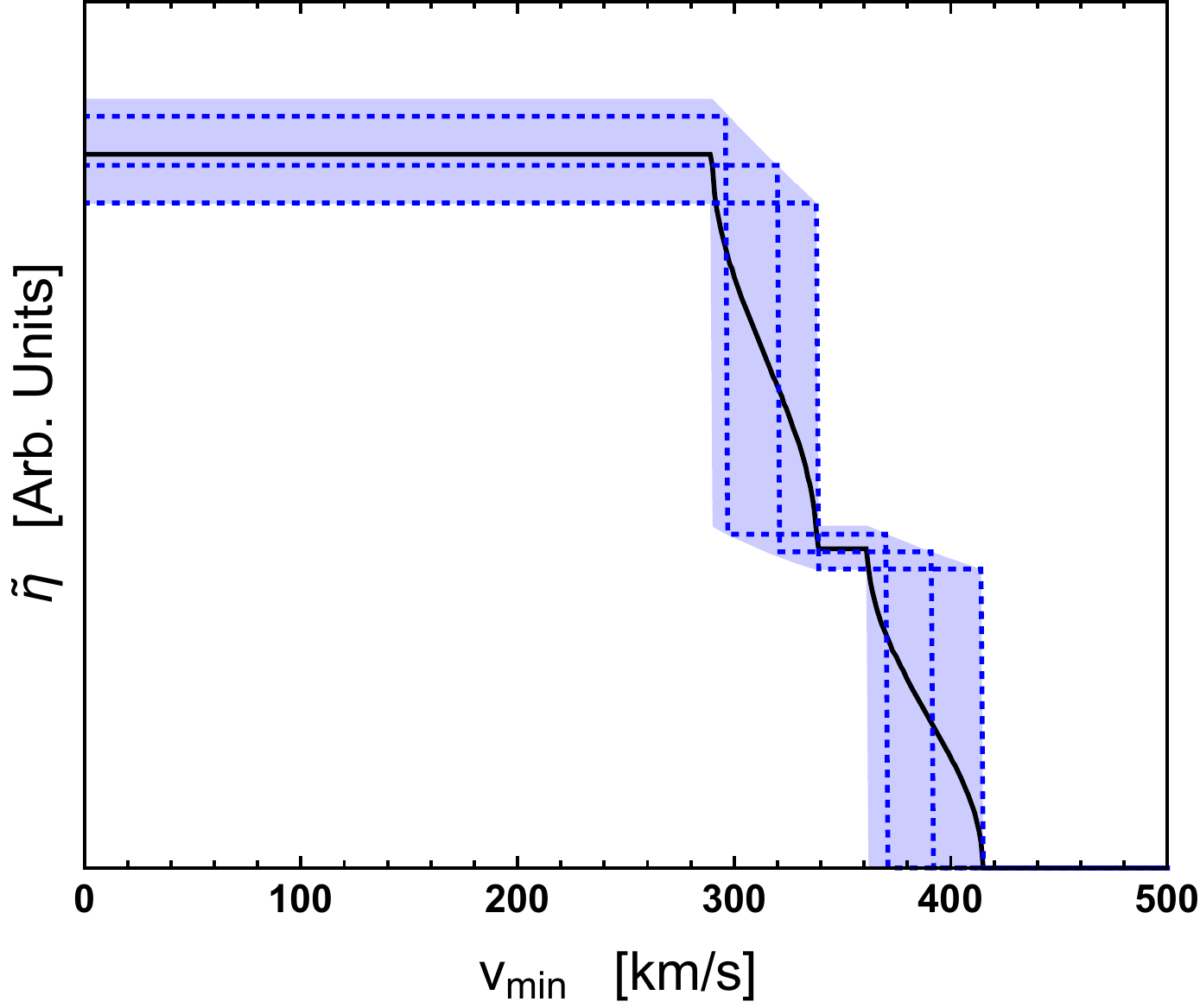}
\includegraphics[width=.49\textwidth]{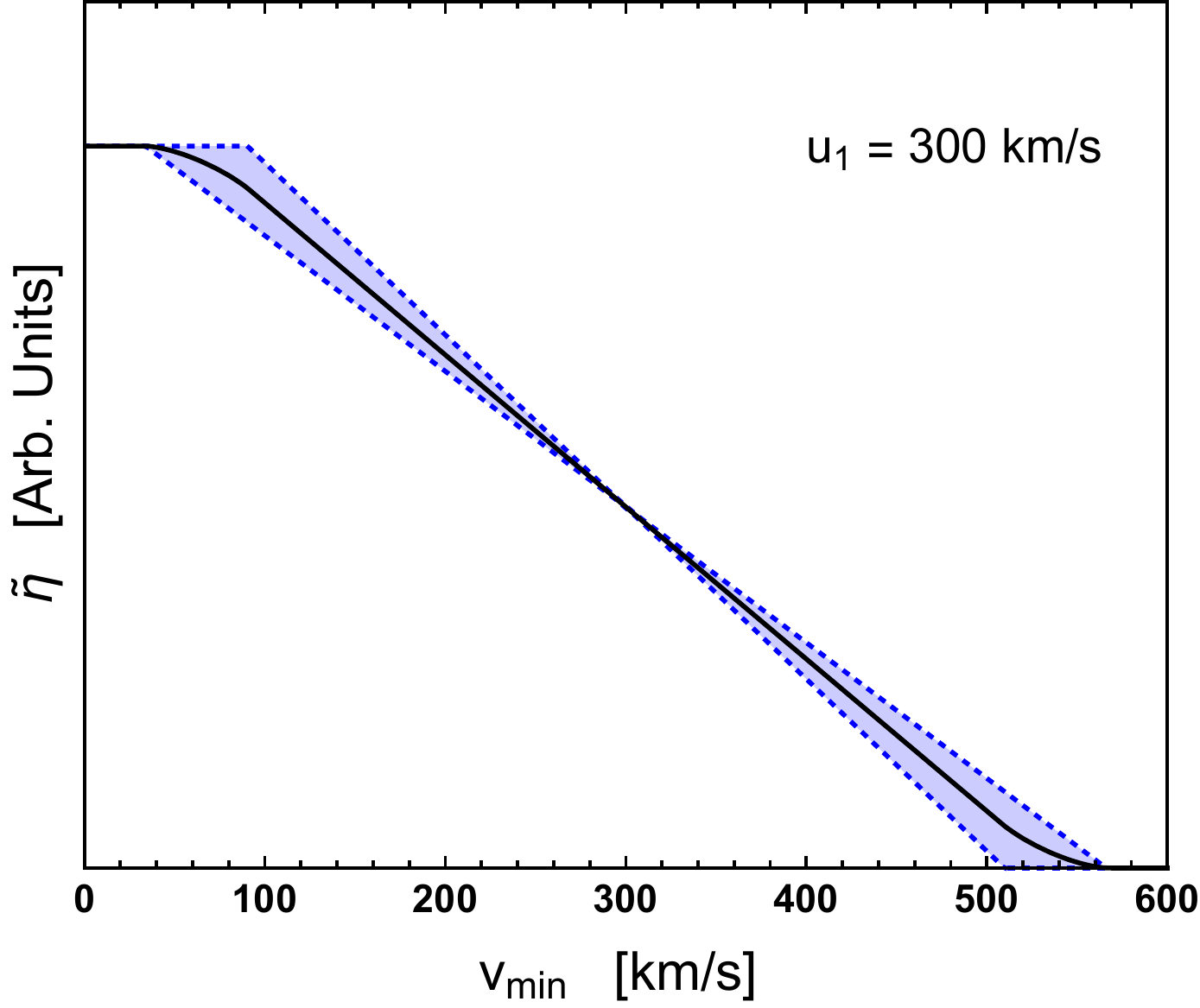}
\caption{\label{fig:etas}
\ref{fig:etas}.a (left) Example of $\teta_{BF}(\vmin,t)$, \Eq{eq:etaBF_t}, for several times (blue dashed lines) for a velocity distribution in \Eq{eq:fgalsum} consisting of two Galactic streams, parameter space spanned in one year by the periodic $\tilde\eta_{BF}(\vmin,t)$ function (blue band), and its time-average $\teta_{BF}^0(\vmin)$,\Eq{eq:etaBF_0} (black line).
\ref{fig:etas}.b (right) The same as in \ref{fig:etas}.a, but for \Eq{eq:etaBF_iso} (isotropic velocity distribution) assuming only one term with $u_1=300$ km/s.
}
\end{figure}

Thus, the formulation presented in this section, based on the Fenchel-Eggleston theorem and assuming a static Galactic DM velocity distribution, can be used in the halo-independent analysis of modulation amplitudes or in a combined analysis of average rates and modulation amplitudes, by combining $\vec R^a$ with $a=0,1,2,\dots$ in a single vector (with dimension given by the sum of the dimension of $\vec R^0$, $\vec R^1$, $\vec R^2$, etc).
With proper modification of some of the equations the same procedures can be used when the transformation from Galactic to Earth's frame is not as simple as the Galilean transformation assumed here.

\section{Summary}\label{sec:summary}

Halo-independent analyses compare direct detection data without the need for astrophysical assumptions. It has been proven in \cite{Fox:2014kua, Gelmini:2015voa, Gelmini:2016pei} using only unmodulated rate data that if the likelihood being analyzed contains at least one extended likelihood, then the best-fit time-averaged halo function $\tilde\eta^0_{BF}(\vmin)$ can always be expressed as a piecewise constant function with a small predetermined number of downward steps. Furthermore, it was shown that this best-fit halo function is unique (see \cite{Gelmini:2016pei}), and could be used to construct two-sided pointwise confidence band in the halo-independent parameter space (see \cite{Gelmini:2015voa, Gelmini:2016pei}). There are two strong limitations to these results: they could not be applied exclusively to binned data and they could not be applied to the analysis of modulation amplitudes. Here we eliminate these two limitations and extend the methods to the analysis of any type of direct detection data.

In Secs.~\ref{Likelihood}--\ref{sec:examples} we extend our halo-independent method to deal with only binned unmodulated rate data. Using theorems on convex hulls we find that
the DM speed distribution in Earth's frame $F(v)$ that maximizes any likelihood can always be expressed as the sum  at most $(\mathcal{N} - 1)$ delta functions in speed, where $\mathcal{N}$ is the total number of data entries (see \Eq{eq:curlyN}).
 Note that this type of best-fit speed distribution yields a piecewise constant best-fit $\tilde\eta$ function $\tilde\eta^0_{BF}$, exactly as we had found before in \cite{Gelmini:2016pei}. However, when using exclusively binned data, this function is not guaranteed to be unique. We thus show how to test the uniqueness of  $\tilde\eta^0_{BF}$.  If it is unique, we show how to produce halo-independent confidence bands at any desired confidence level; if it is not unique, a procedure is given for identifying the degeneracy band, \ie the region of parameter space in which degenerate halo functions which maximize the likelihood reside. 

Finally, in \Sec{sec:modulation} we extend our halo-independent method  to 
the measurements of any coefficient of a harmonic expansion of the time-dependent rate. We work in Galactic coordinates, in which the DM velocity distribution is time-independent and the particle candidate and experiment dependent response functions are time-dependent. We show that the  Galactic velocity distribution $f^{\rm gal}(\vec{u})$ that maximizes any likelihood consists of a sum of at most ${\mathcal N}$
delta functions in Galactic velocity (\ie a sum of Galactic streams with zero velocity dispersion). This type of velocity distribution yields a periodic 
$\tilde\eta_{BF}(\vmin, t)$ which at any fixed time is a piecewise constant function of 
$\vmin$ with at most ${\mathcal N}$ downward steps. In order to compare data in the usual way, we show how to produce a band in $\tilde\eta^0_{BF}(\vmin)$. At this point there are a number of different ways in which one could compare data.  As a possible choice, we show how taking appropriate time-averages  we can construct pointwise confidence and degeneracy bands in  $\tilde\eta_{BF}^0(\vmin)$.  In the last part of \Sec{sec:modulation}  we briefly discuss the consequences of assuming an isotropic Galactic velocity distribution. We show that this choice leads to a Galactic speed distribution $F(u)$ that is once again a sum of delta functions in galactic speed, which produces a time-dependent $\tilde\eta_{BF}(\vmin, t)$, and consequently a time-averaged piecewise linear $\tilde\eta^0_{BF}(\vmin)$, that may differ significantly from those derived without the isotropic assumption.

We believe that the present work is a significant step towards the development of statistically meaningful comparisons of all types of direct detection data, modulated or unmodulated, binned or unbinned,  in a halo-independent manner.

\section*{Acknowledgments} 
 G.G.~acknowledges partial support from the Department of Energy under Award Number DE-SC0009937. JHH is supported by the CERN-Korea fellowship through the National Research Foundation of Korea. SJW is supported by the European Union's Horizon 2020 research and innovation program under the Marie Sk\l{}odowska-Curie grant agreement No. 674896. 

\appendix

\section{Appendix A}
\subsection{Caratheodory Theorem}
If ${\bf y}$ is a vector in the convex hull of a subset $X$ of a vectors space, then there is a set of $n$ vectors ${\bf x_i}$, $i=1,2,\dots,n$ all belonging to $X$, with $n\leq d+1$, where $d$ is the dimension of the convex hull, such that ${\bf y}$ is a {\it convex combination} of the ${\bf x_i}$ vector~\cite{zbMATH02644208} (see also \cite{zbMATH03084780,zbMATH03140980}). A convex combination is a linear combination with real non-negative coefficients $\lambda_i$ which sum to $1$, i.e.
\begin{equation}
{\bf y}=\sum_{i=1}^n\lambda_i{\bf x}_i
\end{equation}
and $\sum_{i=1}^n\lambda_i=1$.

In other words, any vector which is a convex combination of a set of vector $X$ can be expressed as a convex combination of just $n$ of those vectors in $X$, with $n\leq d+1$ (where $d$ is the dimension of the hull). \Fig{fig:caratheodory_discrete} provides two examples of convex hulls (blue regions) formed from a set of vectors ${\bf x}_i$ (black dots) in two dimensions.

\subsection{Fenchel-Eggleston Theorem}
The Fenchel-Eggleston theorem~\cite{zbMATH03084780,zbMATH03140980} strengthens the Caratheodory theorem by taking into account the connectivity of the set of vectors in $X$.

Consider as in the Caratheodory theorem the convex hull of dimension $d$ of a set $X$ of vectors. If $X$ is made of at most $d$ connected components, the upper bound of the number $n$ of vectors in the Caratheodory theorem is tightened to $d$ (instead of $d+1$). \Fig{fig:caratheodory_continuous} provides illustrations of convex hulls (blue regions) formed from either one (Fig.~\ref{fig:caratheodory_continuous}.a) or two (Fig.~\ref{fig:caratheodory_continuous}.b) connected sets of vectors (black lines).  

\begin{figure}
\centering
\includegraphics[width=.49\textwidth]{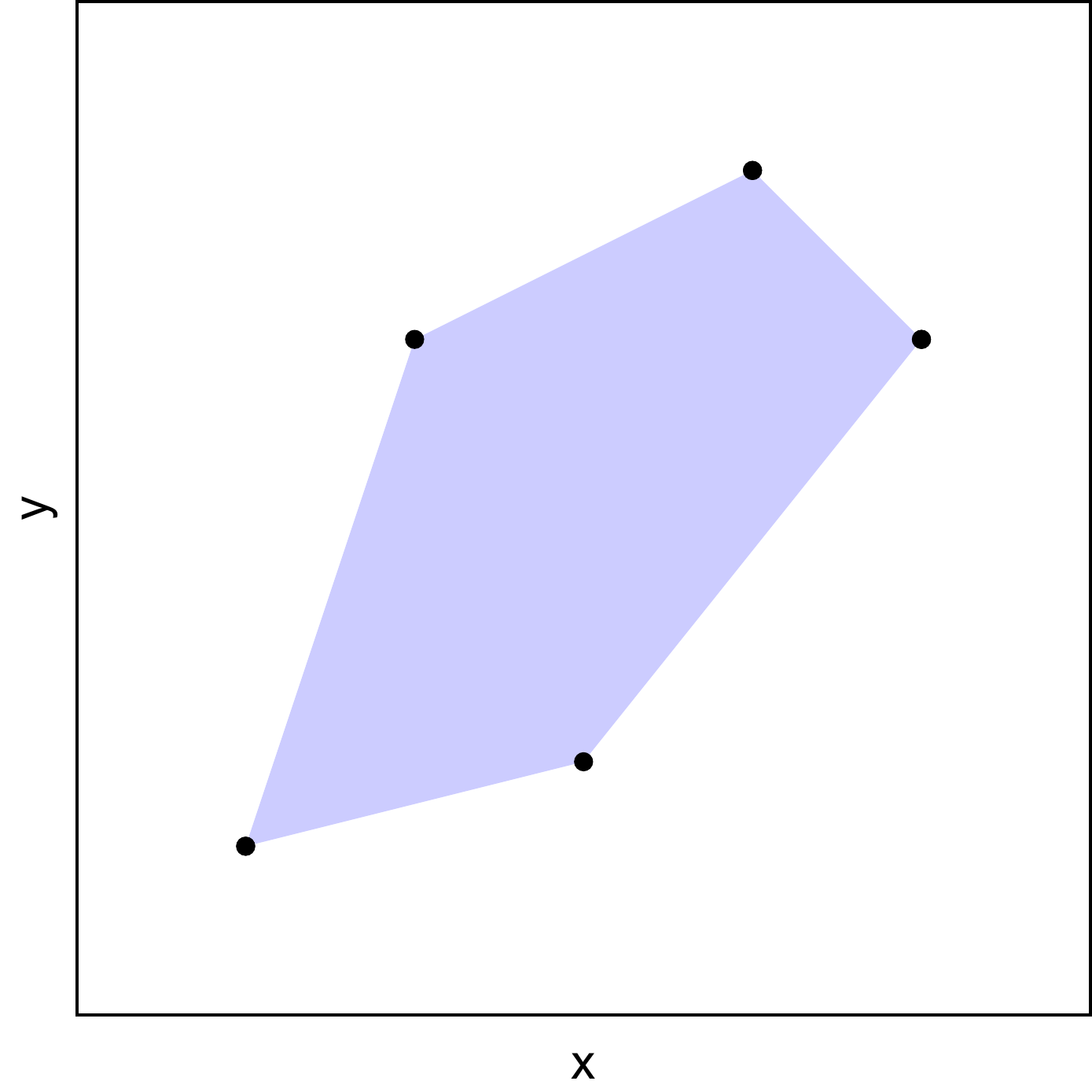}
\includegraphics[width=.49\textwidth]{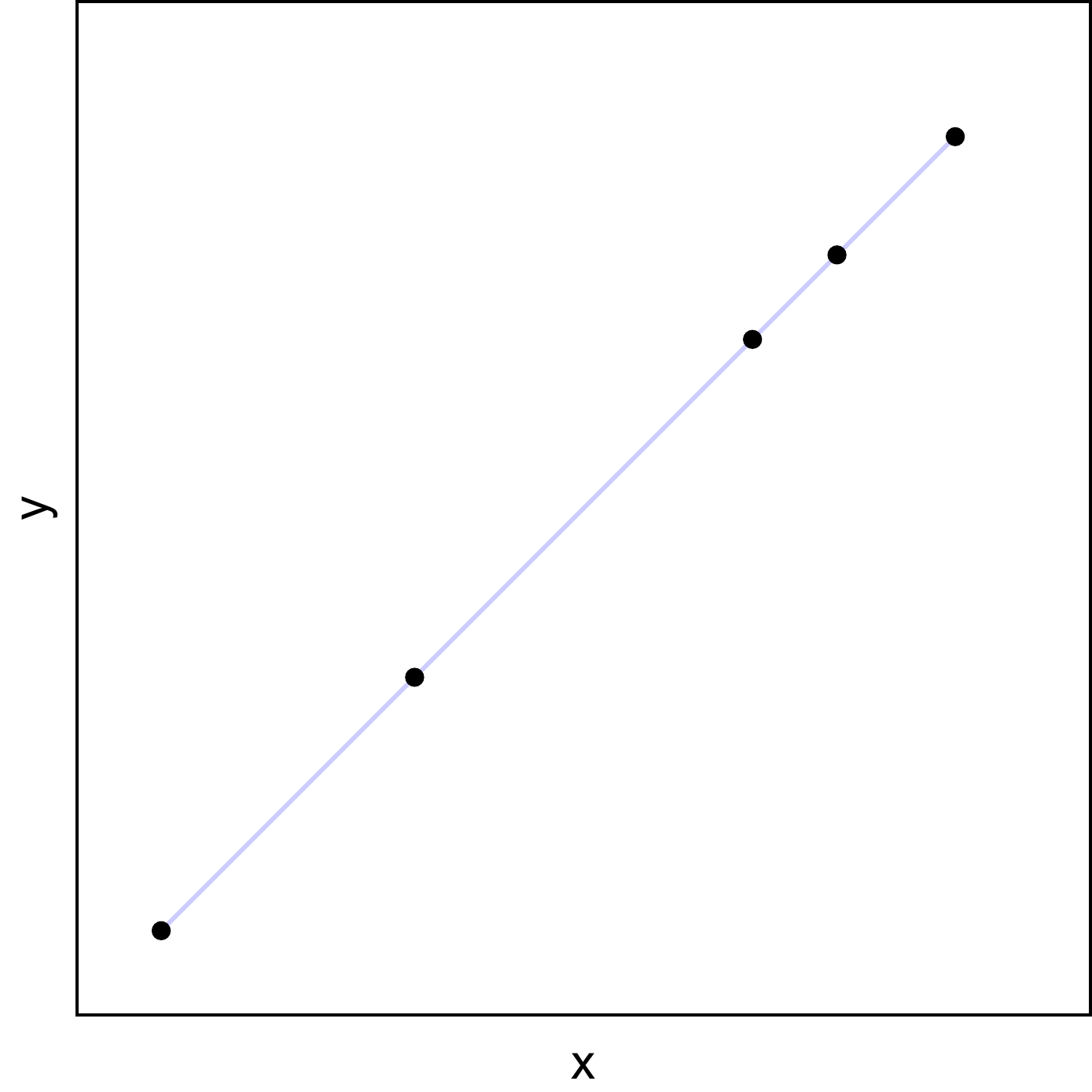}
\caption{\label{fig:caratheodory_discrete} 
\ref{fig:caratheodory_discrete}.a (left) The light blue region is the convex hull of the five points indicated with black dots. Its dimension is $d=2$. The Caratheodory theorem says that any point in the hull can be written as a convex combination of at most $d+1=3$ of the generating points (which is obvious from the figure).
\ref{fig:caratheodory_discrete}.b (right) The convex hull is now the line joining the five generating points in the figure. Its dimension is $d=1$. The Caratheodory theorem says that any one of the points in the line can be written as the convex combination of at most $d+1=2$ points (which is obvious from the figure).}
\end{figure}

\begin{figure}
\centering
\includegraphics[width=.49\textwidth]{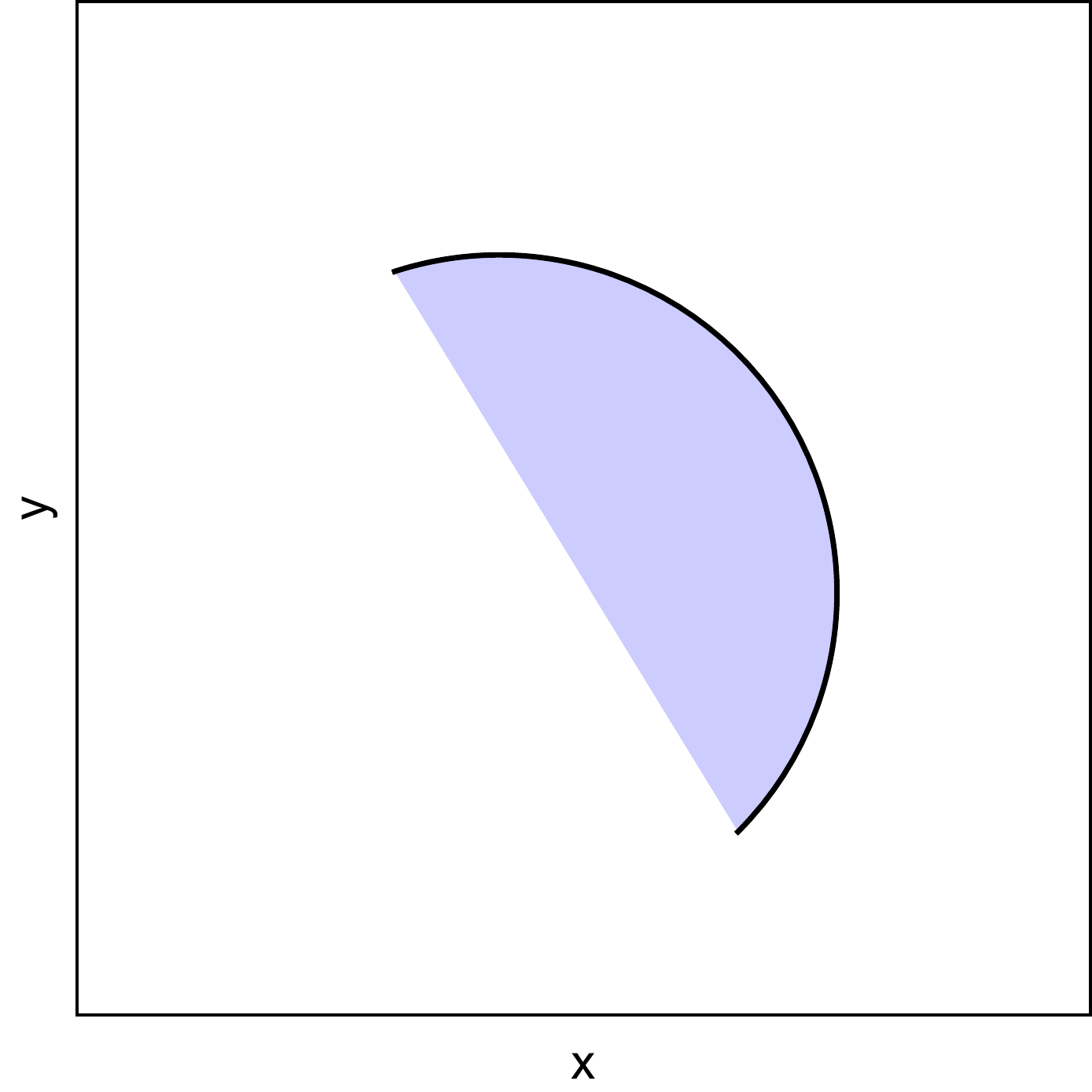}
\includegraphics[width=.49\textwidth]{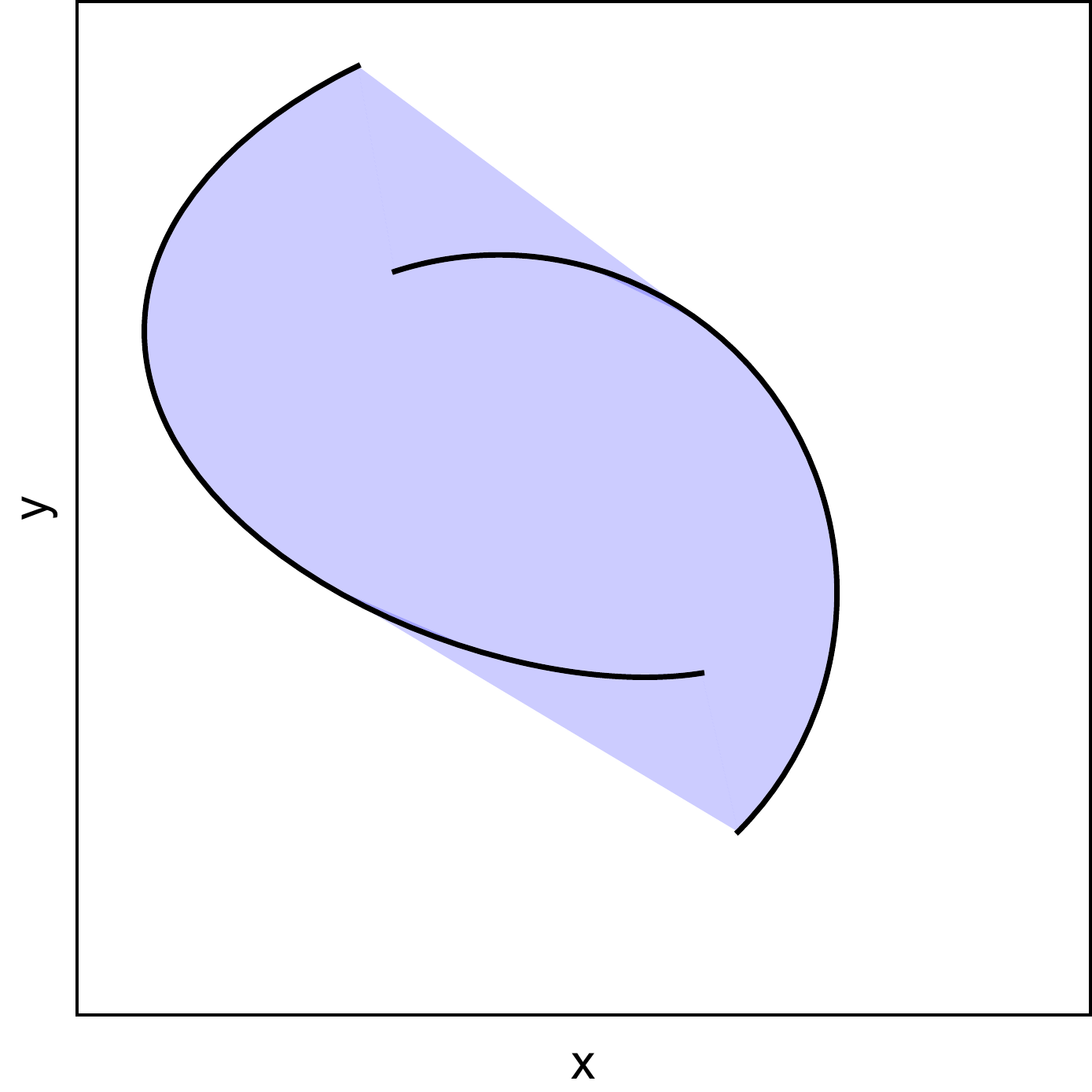}
\caption{\label{fig:caratheodory_continuous} 
 In \ref{fig:caratheodory_continuous}.a (left) there is one line of generating points and in \ref{fig:caratheodory_continuous}.b(right) there are two lines of generating points. In both cases the hulls (light blue regions) have dimension $d=2$ and any point in each of them can be given as the convex combination of at most $d=2$ points of the generating lines. Notice that the theorem applies to a maximum of $d=2$ connected generating set of points (i.e. a maximum of two generating lines in this example).}
\end{figure}

\section{Appendix B}

Assuming the components of any rate vector $\vec R$ sum to a positive number (they are all real non-negative for physically meaningful rates),
\begin{equation}
\sum_{k=1}^{\mathcal N}R_k>0
\end{equation}
we can define the projection of each vector $\vec R$,
\begin{equation}\label{eq:Rhatdef}
\hat R=\frac{\vec R}{\sum_{k=1}^{\mathcal N} R_k}
\end{equation}
on to the plane
\begin{equation}
\sum_{k=1}^{\mathcal N}\hat R_k=1
\end{equation}
which reduces the number of linearly independent components of the $\hat R$ vectors ${\mathcal N}-1$.

We need to define also the projection of the generating vectors $\vec \Hcur(v)$
\begin{equation}\label{eq:Hhat}
\hat\Hcur(v)=\frac{\vec\Hcur(v)}{\sum_{k=1}^{\mathcal N}\Hcur_k(v)},
\end{equation}
onto the same plane
\begin{equation}
\sum_{k=1}^{\mathcal N}\hat\Hcur_k(v)=1,
\end{equation}
which requires
\begin{equation}\label{eq:hatHpositive}
\sum_{k=1}^{\mathcal N}\hat\Hcur_k(v)>0.
\end{equation}

This condition \Eq{eq:hatHpositive} is fulfilled for all $v>0$ for elastic and inelastic exothermic collisions.
However, for inelastic endothermic collisions there may be a minimum speed $(v_\delta)_{min}$ (the smallest of all $v_\delta^T$ speeds defined in \Eq{eq:vdelta} for all components of $\vec\Hcur$) below which all components of $\vec\Hcur$ could be zero, as shown in \Eq{eq:dHcurl}.
In this case we can define $\hat\Hcur(v)$ only for $v>v_{\delta-min}$ (and $v_{\delta-min}=0$ for elastic or inelastic exothermic scattering).

Thus, using the definitions of $\hat R$ (\Eq{eq:Rhatdef}) and $\vec R$ (\Eq{eq:Rvector}) we get
\begin{equation}
\hat R=\frac{\int_{v_{\delta-min}}^\infty{\rm d}v({\mathcal C}\vec\Hcur(v)/v)F(v)}{
\sum_{k=1}^{\mathcal N}\int_{v_{\delta-min}}^\infty{\rm d}u({\mathcal C}\vec\Hcur(u)/u)F(u)}.
\end{equation}
Notice that ${\mathcal C}$ drops out of the definition of $\hat R$.
This easily leads to the definition of all $\hat R$ as the convex hull of $\hat \Hcur(v)$ (defined in \Eq{eq:Hhat})
\begin{equation}
\hat R=\int_{v_{\delta-min}}^\infty{\rm d}v\hat\Hcur(v)\hat F(v)
\end{equation}
where the coefficients $\hat F(v)$ are defined as
\begin{equation}
\hat F(v)=\frac{F(v)\sum_{j=1}^{\mathcal N}\Hcur_j(v)/v}
{\int_{v_{\delta-min}}^\infty{\rm d}u~F(u)\sum_{k=1}^{\mathcal N}\Hcur_k(u)/u}.
\end{equation}
They are real non-negative and it is immediate to verify that,
\begin{equation}
\int_{v_{\delta-min}}^\infty{\rm d}v~\hat F(v) = 1
\end{equation}
as required by the Fenchel-Eggleston theorem. Thus using the theorem we can write any vector $\hat R$ as the convex combination of at most ${\mathcal N}-1$ generating vectors $\hat\Hcur(v_h)$
\begin{equation}\label{eq:Rhatfinite}
\hat R=\sum_{h=1}^{{\mathcal N}-1}\hat\Hcur(v_h)\hat F_h
\end{equation}
with $\sum_{h=1}^{{\mathcal N}-1}\hat F_h=1$, which is equivalent to having
\begin{equation}
\hat F(v)=\sum_{h=1}^{{\mathcal N}-1}\hat F_h\delta(v-v_h).
\end{equation}

We can now prove starting from \Eq{eq:Rhatfinite} that we get to \Eq{eq:Rfinite}, but with $d\leq{\mathcal N}-1$ (instead of $d\leq{\mathcal N}$). Using \Eq{eq:Rhatdef} and \Eq{eq:Rhatfinite} we have
\begin{equation}
\vec R = \hat R \left(\sum_{k=1}^{\mathcal N}R_k\right)
=\left(\sum_{h=1}^{{\mathcal N}-1}\hat\Hcur(v_h)\hat F_h\right) \left(\sum_{k=1}^{\mathcal N}R_k\right)
\end{equation}
Using \Eq{eq:Hhat} we can write
\begin{equation}
\vec R = \sum_{h=1}^{{\mathcal N}-1}\left(\frac{\vec\Hcur(v_h)}{\sum_{j=1}^{\mathcal N}\Hcur_j(v_h)}\right)\left(\sum_{k=1}^{\mathcal N}R_k\right)\hat F_h,
\end{equation}
which coincides with \Eq{eq:Rfinite} but with $d\leq{\mathcal N}-1$ (instead of $d\leq N$) with the identification
\begin{equation}
{\mathcal C}F_h = \hat F_h \frac{v_h\sum_{k=1}^{\mathcal N} R_k}{\sum_{j=1}^{\mathcal N}\Hcur_j(v_h)}\, ,
\end{equation}
and, from the normalization condition $\sum_{h=1}^{{\mathcal N}-1}F_h=1$, we have
\begin{equation}\label{Eq.-mathC}
{\mathcal C} = \sum_{k=1}^{\mathcal N} R_k\sum_{h=1}^{{\mathcal N}-1}\frac{v_h \hat F_h}{\sum_{j=1}^{\mathcal N}\Hcur_j(v_h)}\, .
\end{equation}

For example, in the case of ${\mathcal N}=2$, which corresponds to \Fig{fig:caratheodory_continuous} with the particular choice of ${\mathcal C}$ in \Eq{Eq.-mathC}, the rate vector coincides with one of the generating vector, namely
\begin{equation}
\vec R = \frac{{\mathcal C}\vec\Hcur(v_h)}{v_h}\, ,
\end{equation}
for a particular value of $v_h$. Notice that in this case $F_h=1$.

\bibliographystyle{JHEP}
\bibliography{biblio}

\end{document}